\documentclass[aip,jcp,reprint]{revtex4-1}

\usepackage{amssymb,amsmath}
\usepackage[caption=false]{subfig}
\usepackage{natbib}
\usepackage{url}

\usepackage[mathlines]{lineno}
\usepackage[titletoc,title]{appendix}

\usepackage{graphicx}
\usepackage{hyperref}

\usepackage[usenames,dvipsnames]{color}
\graphicspath{{.},{Figs/},{Figs_eps/}} 
\usepackage{ulem}

\newcommand{\rev}[1]{#1}
\renewcommand{\vec}[1]{\mathbf{#1}}

\newcommand{\rl}{\rho_\textrm{l}}
\newcommand{\rv}{\rho_\textrm{v}}

\newcommand{\tflat}{\Theta_{\textrm{flat}}}
\newcommand{\Pmax}{P^{\rm max}}
\begin{document}

\title{The Cassie-Wenzel transition of fluids on nanostructured
  substrates: Macroscopic force balance versus microscopic
  density-functional theory}

\author {Nikita Tretyakov}
\email{tretyakov@mpip-mainz.mpg.de}
\affiliation{Max-Planck-Institut f\"ur Polymerforschung,
  Ackermannweg 10, 55128 Mainz, Germany}

\author{Periklis Papadopoulos} 
\affiliation{Max-Planck-Institut f\"ur Polymerforschung,
  Ackermannweg 10, 55128 Mainz, Germany}
\affiliation{Physics Department, University of Ioannina,
  P.O. Box 1186, GR-45110 Ioannina, Greece}

\author{Doris Vollmer} 
\affiliation{Max-Planck-Institut f\"ur Polymerforschung,
  Ackermannweg 10, 55128 Mainz, Germany}

\author{Hans-J\"urgen Butt} 
\email{butt@mpip-mainz.mpg.de}
\affiliation{Max-Planck-Institut f\"ur Polymerforschung,
  Ackermannweg 10, 55128 Mainz, Germany}

\author{Burkhard D\"unweg}
\email{duenweg@mpip-mainz.mpg.de}
\thanks{B.~D. and K.~C.~D. contributed equally.}
\affiliation{Max-Planck-Institut f\"ur Polymerforschung,
  Ackermannweg 10, 55128 Mainz, Germany}
\affiliation{Institut f\"ur Festk\"orperphysik, Technische
  Universit\"at, Hochschulstra{\ss}e 12, 64289 Darmstadt, Germany}
\affiliation{Department of Chemical Engineering, Monash University,
  Clayton, Victoria 3800, Australia}

\author {Kostas Ch. Daoulas}
\email{daoulas@mpip-mainz.mpg.de}
\thanks{B.~D. and K.~C.~D. contributed equally.}
\affiliation{Max-Planck-Institut f\"ur Polymerforschung,
  Ackermannweg 10, 55128 Mainz, Germany}

\date{\today}

\begin{abstract}
  Classical density functional theory is applied to investigate the
  validity of a phenomenological force-balance description of the
  stability of the Cassie state of liquids on substrates with
  nanoscale corrugation. A bulk free-energy functional of third order
  in local density is combined with a square-gradient term, describing
  the liquid-vapor interface. The bulk free energy is parameterized to
  reproduce the liquid density and the compressibility of water. The
  square-gradient term is adjusted to model the width of the
  water-vapor interface. The substrate is modeled by an external
  potential, based upon Lennard-Jones interactions. The
  three-dimensional calculation focuses on substrates patterned with
  nanostripes and square-shaped nanopillars. Using both the
  force-balance relation and density-functional theory, we locate the
  Cassie-to-Wenzel transition as a function of the corrugation
  parameters. We demonstrate that the force-balance relation gives a
  qualitatively reasonable description of the transition even on the
  nanoscale. \rev{The force balance utilizes an effective contact
    angle between the fluid and the vertical wall of the corrugation
    to parameterize the impalement pressure. This effective angle is
    found to have values smaller than the Young contact angle.  This
    observation corresponds to an impalement pressure that is smaller
    than the value predicted by macroscopic theory. Therefore, this
    effective angle embodies effects specific to nanoscopically
    corrugated surfaces, including the finite range of the
    liquid-solid potential (which has both repulsive and attractive
    parts), line tension, and the finite interface thickness.}
  Consistently with this picture, both patterns (stripes and pillars)
  yield the same effective contact angles for large periods of
  corrugation.
\end{abstract} 

\maketitle

\section{Introduction}
\label{sec:intro}

\begin{figure*}[t]
  \subfloat[]{\label{fig:cassie}\includegraphics[width=5.4cm]%
    {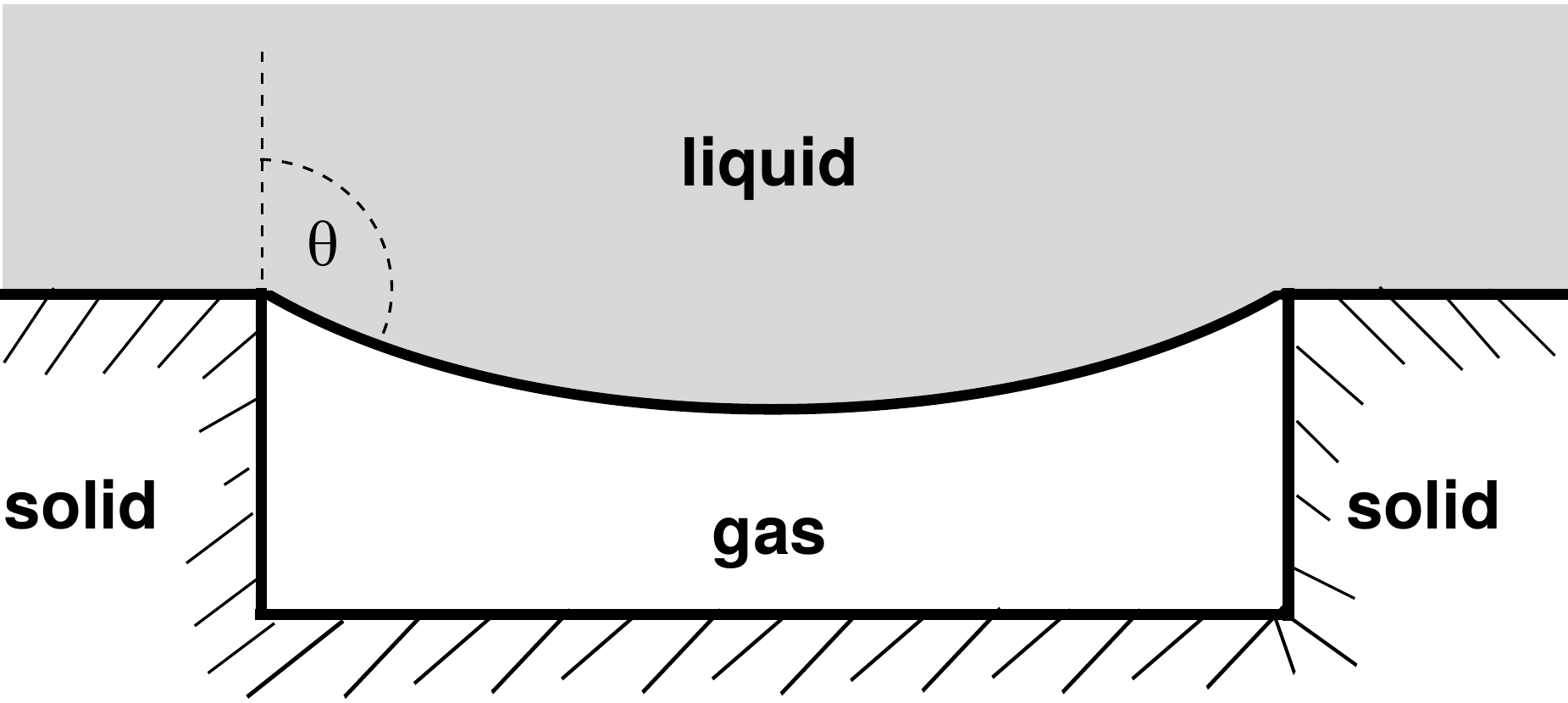}}\hspace*{0.35cm}
  \subfloat[]{\label{fig:wenzel}\includegraphics[width=5.4cm]%
    {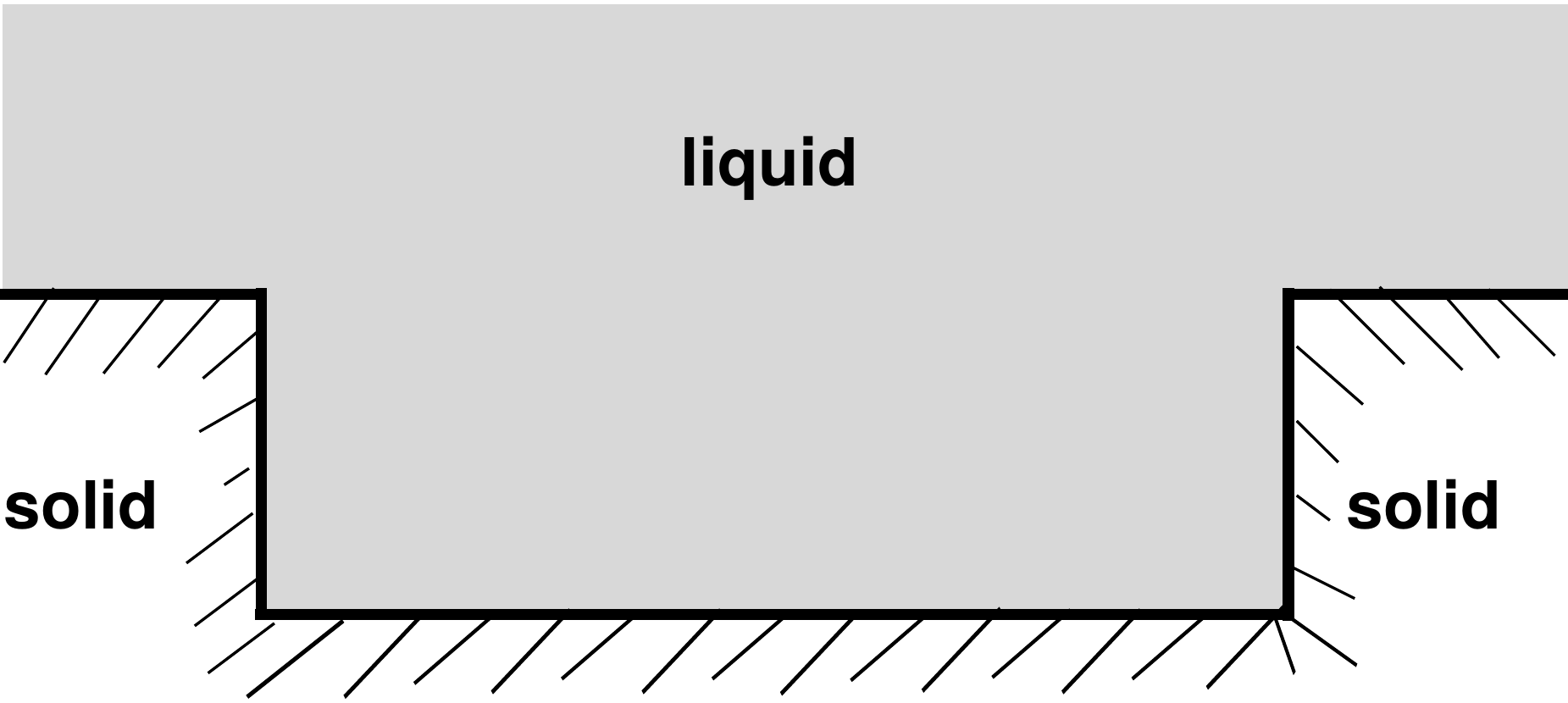}}\hspace*{0.35cm}
  \subfloat[]{\label{fig:cassie_sag}\includegraphics[width=5.4cm]%
    {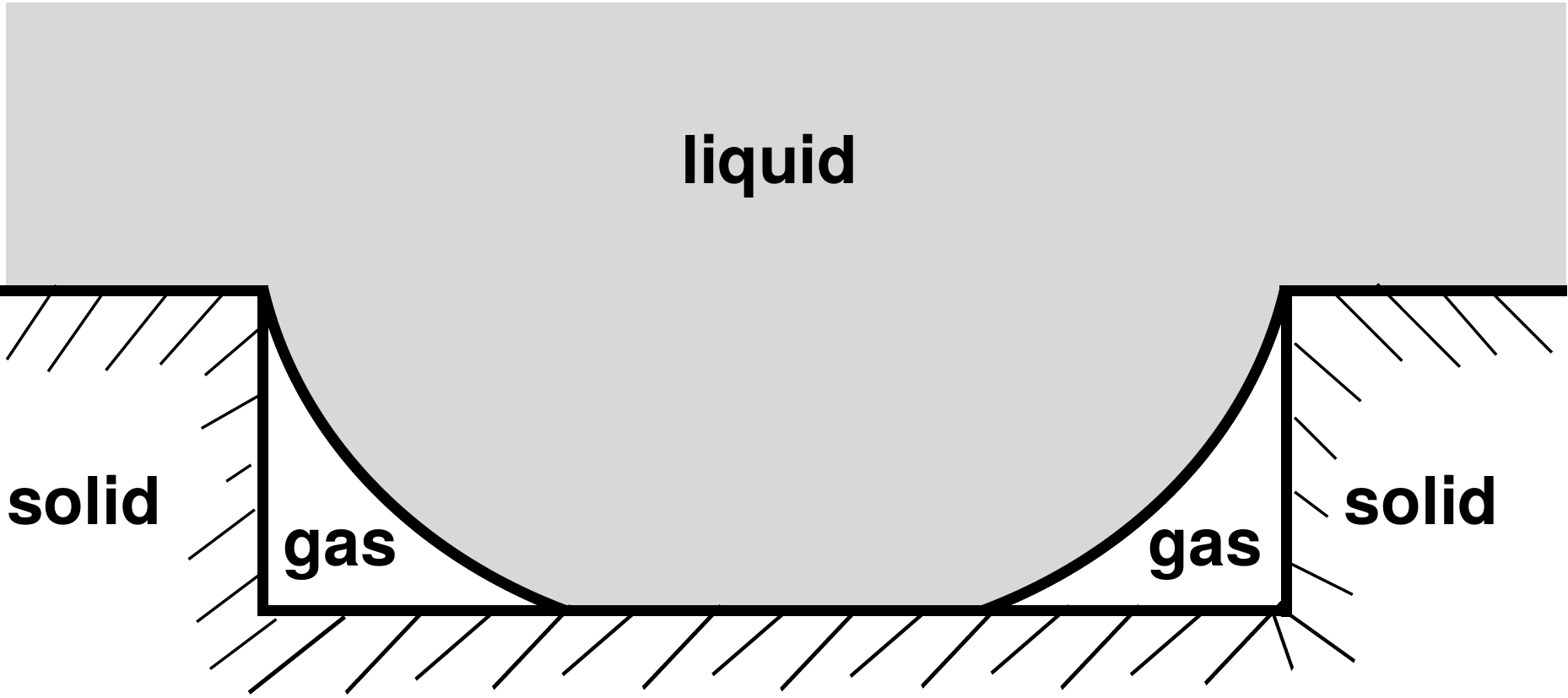}}
  \caption{Various wetting regimes in the simplified two-dimensional
    geometry: (a) Cassie state, where the liquid-gas interface resides
    on top of the solid corrugations; (b) Wenzel state, where the
    liquid permeates the cavity between the topographic structure, and
    (c) Sagging mechanism in the Cassie state, where the liquid-gas
    interface still resides on top of the corrugations but due to its
    curvature (or corrugation dimensions) gets into contact with the
    bottom of the cavity.
  \label{fig:states}
  }
\end{figure*}

Super liquid-repellent surfaces are important for fundamental studies
of wetting phenomena, as well as technological applications, including
anti-fouling coatings~\cite{Antifouling}, fog harvesting~\cite{Fog},
drag reduction~\cite{Watanabe, Gogte, Drappier, JO_BP_JR_2004}, and
gas exchange membranes~\cite{McHale, MPIP}. Micro- and nanoscale
topographic structuring offers significant opportunities for
manufacturing such surfaces, by introducing a special wetting mode --
the Cassie or the ``fakir'' state~\cite{AC_SB_1944}.  In this regime,
the liquid resides on top of the topographic corrugations, while the
enclosed space below the liquid is filled with gas
(Fig.~\ref{fig:cassie}). This reduces the adhesion, and a droplet can
roll off easily. Thermodynamically, however, the Cassie state is
frequently only metastable~\cite{BormaRev, CalliesQuere, DorrerRuhe,
  Nosonovsky}, meaning that it corresponds to only a local minimum of
the free energy, compared to the free energy of a slightly deformed
droplet. In such a metastable situation, the absolute free-energy
minimum rather corresponds to the complete-wetting or Wenzel
state~\cite{Wenzel_1936}. In the Wenzel state, the liquid permeates
the topographic structure, such that no gas ``pockets'' remain, the
adhesion is increased, and the contact angle exhibits a large
hysteresis (Fig.~\ref{fig:wenzel}). The Wenzel state is therefore
usually not super-repellent~\cite{CalliesQuere}, and hence most
technological applications aim at avoiding it as much as possible.

To understand the basic physics of the system, let us consider the
coexistence of the liquid and the gas phase, in the presence of the
surface. Let us also assume, as a first step, that the surface is
flat. Since it is repelling, it is clear that an open system at a
pressure slightly above the bulk coexistence pressure will be
``dry''. In other words, the condensed liquid phase exists away from
the surface, while direct contact with it is avoided. However, the
``dry'' layer is very thin; its thickness is roughly comparable to the
range of interaction of the repulsion. From a macroscopic point of
view, this means contact, but with a large interfacial tension
$\gamma_{\rm sl}$ between solid and liquid. Now we add a groove to the
surface, and we first assume that it is infinitely deep. In this
situation, the system can either be in the Cassie state, where the
liquid-gas interface remains suspended above the groove, or in the
Wenzel state, where the whole groove is filled with liquid. This is so
because pushing the interface into the groove by a length $h$ results
in a gain of bulk free energy that scales linearly with $h$, but also
in a loss of liquid-solid surface free energy that scales linearly
with $h$ as well.  Since the system prefers a liquid-gas interface
over a liquid-solid interface, the Cassie state is preferred, unless
the bulk pressure is so large that the bulk free energy prevails and
the system chooses the Wenzel state. For a groove that is an
infinitely long stripe, this consideration results in a critical
excess pressure (additional pressure above the gas value) of
\begin{equation}
  \Pmax = - \frac{2 \gamma}{D} \cos \tflat ,
\end{equation}
where $D$ is the width of the groove, $\gamma$ the liquid-gas
interface tension, and $\tflat$ the Young contact angle of
a macroscopic liquid droplet on a flat surface (larger than
$90^\circ$, due to the repelling surface). For a derivaton of this
relation~\cite{BormaRev, DorrerRuhe, Extrand1, Zheng, Bartolo,
  Bizonne, Moulinet, HJB_DV_PP_2014}, see Appendix~\ref{sec:appendix}.
Within this picture, the Cassie-Wenzel transition may therefore be
viewed as a first-order phase transition. From the experimental point
of view, it is useful to re-write the relation as
\begin{equation}
  \Pmax = - \frac{L_{\rm c}}{A}
  \frac{\gamma \cos \Theta_{\rm max}}{1 - f} ,
  \label{eq:hj_gen}
\end{equation}
where we assume that the total surface has an area $A$, of which a
fraction $f$ is wetted in the Cassie state, while the fraction $1 - f$
is covered with gas. More precisely, we mean by $A$ the area that is
obtained after projecting the nanostructure onto the surface, and
similarly for the fractions $A f$ and $A (1 - f)$. Furthermore,
$L_{\rm c}$ denotes the contour length of the contact line, while
$\Theta_{\rm max}$ replaces $\tflat$ as the maximum contact angle;
this is an effective angle that takes into account that the grooves in
general have a different shape than one-dimensional
stripes~\cite{Moulinet}. It should also be noted that the physical
origin of the pressure is of course arbitrary~\cite{Bartolo}; e.~g. it
can be the Laplace pressure due to the curvature of the deposited
droplet, a dynamic pressure during droplet impact on the surface, or
hydrostatic pressure (underwater
superhydrophobicity~\cite{underwaterHemeda}).

This simple phenomenological picture has been experimentally
confirmed, e.~g., in the observation of evaporating
droplets~\cite{Moulinet}. Here the Laplace
pressure~\cite{deGennesBook} keeps on increasing during the
evaporation process, until it reaches $\Pmax$ (and the contact angle
$\Theta_{\rm max}$), and the Cassie state collapses.

However, in general we may have to take into account that the groove
has a finite depth. In this situation, the Cassie and Wenzel states
look like depicted in Figs.~\ref{fig:cassie} and~\ref{fig:wenzel},
respectively. Apart from these two cases, it is then in principle
possible that yet a third stable state occurs. It is depicted in
Fig.~\ref{fig:cassie_sag} and we refer to it as the ``sagged Cassie''
(SC) state. Here the contact line at the ``upper'' end of the groove
still remains pinned, but the liquid-gas interface is already in
contact with the ``bottom'' of the cavity, while gas pockets remain in
the corners of the groove. Such a state typically requires strong
hydrophobicity, in order to make sure that the contact angle at the
onset of sagging is still below the Young angle. Another possibility
to reach the SC state consists of very shallow grooves, with a small
depth-to-width ratio. However, for these latter systems the involved
pressures are quite small, which means that they are very difficult to
control and handle. For these reasons, the SC state has so far rarely
been of experimental importance. Conversely, within the framework of
theoretical calculations as described below, it is, for large
hydrophobicity, quite easily possible to observe the SC state. As a
matter of fact, we did observe it for a system with a Young contact
angle of roughly $150^\circ$. However, the purpose of the present
paper is mainly a test of the theory of the Cassie-Wenzel transition
(i.~e. essentially Eq.~(\ref{eq:hj_gen})), which obviously does not
apply to the SC state. For this reason, we have deferred a study of
these phenomena to future work, and confined the present investigation
to systems where a direct Cassie-Wenzel transition occurs.

\rev{Equation~(\ref{eq:hj_gen}) shows clearly the challenges of
  manufacturing good super-repellent surfaces. Firstly, one should
  realize that the ratio $L_{\rm c} / A$ is $\sim l^{-1}$, where $l$
  is the characteristic length scale of the corrugation geometry,
  assuming that its basic structure type (stripes, pillars, etc.)
  remains fixed. Now, one would like to make $f$ as small as possible
  in order to facilitate easy roll-off~\cite{CalliesQuere}. However,
  to stabilize the Cassie state (or to increase $\Pmax$) one would
  like to make $f$ as large as possible. In principle, this problem is
  avoided if $l$ is decreased at constant $f$, meaning that both the
  period of the corrugation and the lateral dimensions of the
  asperities are decreased at the same rate: In this case, $\Pmax$
  increases without sacrificing the frictional properties. However,
  this ``miniaturization route''~\cite{Extrand2, Zheng,
    Moulinet,BormaRev, HJB_DV_PP_2014} is of course experimentally
  difficult.}

It is thus obvious that it is highly interesting to understand the
Cassie-Wenzel transition for nanoscopically structured surfaces. Here,
however, one should take into account that Eq.~(\ref{eq:hj_gen}) is
based upon a \textit{macroscopic} consideration. At the nanoscale,
there are many aspects beyond macroscopic physics, which potentially
can alter the picture -- or at least modify it when it comes to
quantitative considerations. Such aspects are (i) the finite thickness
of the liquid-gas interface, (ii) line tension effects, (iii) the
finite range of the substrate potential, (iv) the atomic structure of
the substrate, (v) the atomic structure of the liquid, and (vi)
thermal fluctuations. It is thus not quite clear to what extent
Eq.~(\ref{eq:hj_gen}) remains valid at the
nanoscale~\cite{HJB_DV_PP_2014}. Recent {\it in situ} x-ray
diffraction studies~\cite{DietrichPRL} of the Cassie-Wenzel transition
indicate that this might be indeed the case. However, to the best of
our knowledge, extensive theoretical studies of this question are
still lacking, and the present paper is intended as a step in this
direction. We therefore investigate a simple microscopic model and
compare its predictions for the stability of the Cassie state with
Eq.~(\ref{eq:hj_gen}). In the present study only the microscopic
aspects (i)--(iii) are taken into account, such that the calculations
are not too compute-intensive, and a large range of parameters can be
explored. It is possible to refine the model and include the aspects
(iv)--(vi); this is however left for future work.

Similar theoretical studies exist, which investigate microscopic models at 
structured surfaces. However, these focus mainly on the dynamics, which 
is of course also highly interesting and important. Apart from analytical 
theories~\cite{Marmur2, Patankar} there is currently significant interest in 
exploring the barriers and the transition pathways between the two states. This 
can be done by lattice Boltzmann simulations~\cite{Yeomans, Varnik}, 
string-method calculations cast in the framework of continuum 
descriptions~\cite{Ren} and particle-based Molecular Dynamics 
(MD)~\cite{Giacomello}, as well as Boxed MD~\cite{Escobedo}.

In the current study, we employ an approach based on the classical
density functional theory (DFT) for liquids. Due to the flexibility of
the formalism, wetting phenomena can be addressed within DFT at
different levels of detail, as has been extensively discussed in
several reviews~\cite{Evans, Dietrich, Tarazona1, Wu, Weeks}. A simple
strategy, which we employ here, is to consider functionals which
neglect the molecular structure of the liquid and therefore do not
account for, e.~g., layering at walls. At this level, the interactions
between liquid molecules (i.~e. the fact that below the liquid-gas
transition the molecules tend to aggregate) is just taken into account
by the square-gradient approximation~\cite{CahnHilliard}, while the
bulk thermodynamics is described by a strictly local functional.

More sophisticated functionals, like, e.~g., the weighted density
approximation~\cite{Tarazona2}, or even more elaborate
schemes~\cite{Lum, Hughes, Sullivan1, Sullivan2, Haslam, Lischner,
  Sundararaman}, can be constructed, such that details of short-range
correlations in the liquid are faithfully modeled. However, these
elaborate functionals are computationally much more expensive, and
hence have not been used in the present exploratory
study. Representative studies of liquids on chemically or
topographically patterned substrates (but only in two dimensions!),
based upon such functionals, can be found in Refs.~\cite{Koch,
  Malijevsky, Berim}, while our simple functional allows us to study
three-dimensional systems without major problems. This DFT model is
parameterized on the basis of a few thermodynamic properties of water,
known from experiments. The liquid-substrate interactions are taken
into account by a Lennard-Jones potential. Several cases of strength
of water/solid interactions are considered, realizing different
degrees of hydrophobicity of the surface. \rev{A highly related
  previous DFT study by Zhang and Ren~\cite{ZhangRen} was done in
  three dimensions as well, and resulted in Cassie-Wenzel phase
  diagrams for various surface patterns. In contrast to the present
  study, whose aim is the comparison with macroscopic theory,
  Ref.~\onlinecite{ZhangRen} rather focused on transition states
  investigated with the ``string method'' and did not attempt to
  adjust the functional to an experimental system. Our study should
  therefore be viewed as complementary to Ref.~\onlinecite{ZhangRen}.}

The paper is organized as follows: Sec.~\ref{sec:models} presents the
modeling strategy. In particular, the classical DFT approach is
introduced in Sec.~\ref{ssec:dft}, while Sec.~\ref{ssec:param}
explains the parameterization. The method of incorporating corrugated
substrates into the model is discussed in Sec.~\ref{ssec:subs} and the
numerical scheme is elaborated in Sec.~\ref{ssec:numscheme}. We
proceed with contact angle calculations in the canonical ensemble in
Sec.~\ref{ssec:canonical}. The predictions of the phenomenological
force balance and the classical DFT calculations on nanostructured
substrates are compared in the remainder of
Sec.~\ref{sec:water-vapor}. We conclude in Sec.~\ref{sec:summ} with a
short summary and an outlook.

\section{Modeling approach}
\label{sec:models}

\subsection{Density functional theory description}
\label{ssec:dft}

We study the equilibrium thermodynamics of a system with volume $V$ and 
temperature $T$ in the grand-canonical ensemble. The starting point is a 
grand-canonical potential $\Omega$, which is a functional of the average local 
number density $\rho({\bf r})$:
\begin{eqnarray}
  \Omega
  & = & 
  \int d{{\bf r}} \; \omega (\rho(\vec{r}) , \nabla \rho(\vec{r})) ,
  \\
  \frac{\omega}{k_BT}
  & = & 
  \rho({\bf r}) \left[ \ln( \Lambda^3 \rho({\bf r})) - 1 \right]
  + f_{\rm b}(\rho({\bf r})) + \nonumber \\
  & & 
  \frac{\kappa}{2} \left( \nabla \rho({\bf r}) \right)^2
  + U_\textrm{s}({\bf r}) \rho({\bf r}) - \mu \rho({\bf r}) .
  \label{eq:grandcanfree}
\end{eqnarray}

Here $k_BT$ is the thermal energy, such that $\Omega / k_B T$ is
dimensionless.  The first term in Eq.~(\ref{eq:grandcanfree})
describes the translational entropy of the liquid molecules. $\Lambda$
is usually taken as the thermal de Broglie wavelength, which gives
rise to a normalization volume $\Lambda^3$ that is needed for
dimensional reasons. Actually, the value of $\Lambda$ is immaterial,
since it only serves to define the zero of the dimensionless chemical
potential $\mu$ in the last term. This is seen from the trivial
identity
\begin{equation}
  \rho \ln(\Lambda^3 \rho) - \mu \rho =
  \rho \ln\left( \frac{\rho}{\Lambda^{-3} \exp(\mu)} \right) ,
\end{equation}
which means that the only relevant parameter is the combination
$\Lambda^{-3} \exp(\mu)$. The bulk excess Helmholtz free energy per
unit volume and per $k_B T$ is denoted by $f_{\rm b}(\rho({\bf
  r}))$. The square-gradient term penalizes the presence of liquid-gas
interfaces~\cite{CahnHilliard}. $\kappa$ controls the interfacial
width and, in general, can be density-dependent~\cite{Yang1976,
  Anderson}. In the present study, $\kappa$ is assumed to be a
constant. The influence of the substrate is described by an external
potential (per $k_B T$) denoted by $U_{\rm s}({\bf r})$. The form of
this potential, specific to the current study, is discussed in
sec.~\ref{ssec:subs}. It should be noted that $U_{\rm s}$ is
dimensionless, while $f_{\rm b}$ has the dimension of an inverse
volume. For $f_{\rm b}(\rho({\bf r}))$ we choose
\begin{equation}
  f_{\rm b}(\rho({\bf r})) = \frac{v}{2}\rho({\bf r})^2
  + \frac{w}{3}\rho({\bf r})^3
  \label{eq:Helmbulk}
\end{equation}
with constant coefficients $v$ and $w$. This is a very simple model
capable of describing liquid-gas coexistence, due to the competition
between attraction ($v < 0$) and repulsion ($w > 0$). These
coefficients, and also the interface parameter $\kappa$, are chosen to
reproduce some reference properties of liquid water (see
Sec.~\ref{ssec:param} for details).

The equilibrium density distribution minimizes $\Omega$ and therefore
fulfills the corresponding Euler-Lagrange equation:
\begin{equation}
  \ln(\Lambda^3 \rho({\bf r})) + 
  \frac{\partial f_{\rm b}(\rho({\bf r}))}{\partial \rho({\bf r})} 
  -\kappa \vec{\nabla}^2 \rho({\bf r}) + U_\textrm{s}({\bf r}) - \mu = 0 .
  \label{eq:EulLangrangeGC}
\end{equation}

To re-cast this DFT description in terms of Self Consistent Field
(SCF) theory~\cite{Thomson, MullerJCP2003, BrykMacDowell}, it is
instructive to rewrite Eq.~(\ref{eq:EulLangrangeGC}) as a system of
equations:
\begin{eqnarray}
  W({\bf r}) & = & 
  \frac{\partial f_{\rm b}(\rho({\bf r}))}{\partial \rho({\bf r})} 
  -\kappa \vec{\nabla}^2 \rho({\bf r}) + U_\textrm{s}({\bf r}) , 
  \label{eq:SCF_GC1} \\
  \rho({\bf r}) & = & \Lambda^{-3} \exp(\mu)\exp(- W({\bf r})) .
  \label{eq:SCF_GC2}
\end{eqnarray}
 
The quantity $W({\bf r})$ plays the role of a mean field representing
the interactions of a liquid molecule with its surrounding molecules
and the substrate. Apart from illustrating the link to the SCF theory
framework, Eqs.~(\ref{eq:SCF_GC1}) and~(\ref{eq:SCF_GC2}) also provide
the basis for our numerical scheme.

For a bulk liquid with density $\rho_{\rm b}$, the density profile is
constant, and the surface potential vanishes. In this situation,
Eqs.~(\ref{eq:SCF_GC1}) and~(\ref{eq:SCF_GC2}) are simplified to
\begin{equation}
  \Lambda^{-3} \exp(\mu) = \rho_{\rm b}
  \exp\left(\frac{\partial f_{\rm b}}{\partial \rho}
  \Big \vert_{\rho = \rho_{\rm b}}\right) .
  \label{eq:chem_pot_def}
\end{equation}

Inserting Eq.~(\ref{eq:chem_pot_def}) into Eq.~(\ref{eq:SCF_GC2}), one
sees that instead of the chemical potential we can rather use
$\rho_{\rm b}$ in the liquid phase as a control parameter.

For our nanoscopic system with an interface, the concept of pressure
needs to be generalized to the concept of a stress tensor $\tensor \Pi
(\vec{r})$. From the general principles of Lagrangian field
theory~\cite{LL_EL_75} it is clear that the stress tensor is
calculated as
\begin{equation}
  \tensor \Pi = \frac{\partial \omega}{\partial \nabla \rho}
  \otimes \nabla \rho - \omega \; \tensor 1 ;
\end{equation}
for our functional this results in~\cite{Yang1976}
\begin{eqnarray}
  \frac{\tensor \Pi}{k_B T}
  & = &
  \kappa \left[ \nabla \rho \otimes \nabla \rho
    - \frac{1}{2} \left( \nabla \rho \right)^2 \tensor 1 \right]
  \\
  & &
  - \left[ \rho ( \ln (\Lambda^3 \rho) - 1) + f_{\rm b}
    + U_s \rho - \mu \rho \right] \tensor 1 .
  \nonumber
\end{eqnarray}

For an infinite homogeneous bulk system, the expression becomes much simpler: 
$\tensor \Pi = P \tensor 1$ with
\begin{equation}
  \frac{P}{k_B T} = - \rho_{\rm b}(\ln(\Lambda^3 \rho_{\rm b}) - 1) -
  f_{\rm b}(\rho_{\rm b}) + \mu \rho_{\rm b} .
\end{equation}

Again we can use Eq.~(\ref{eq:chem_pot_def}) to eliminate the chemical
potential; this results in the equation of state
\begin{equation}
  \frac{P}{k_B T} = \rho_{\rm b} +
  \rho_{\rm b} \left. \frac{\partial f_{\rm b}}{\partial \rho}
  \right\vert_{\rho = \rho_{\rm b}} - f_{\rm b}(\rho_{\rm b}) .
\label{eq:pressure_in_bulk}
\end{equation}

The stress tensor can therefore also be re-written as
\begin{eqnarray}
  \frac{\tensor \Pi}{k_B T}
  & = &
  \kappa \left[ \nabla \rho \otimes \nabla \rho
    - \frac{1}{2} \left( \nabla \rho \right)^2 \tensor 1 \right]
  \nonumber
  \\
  &&
  + P_{\rm bulk} \tensor 1 - U_{\rm s} \rho \tensor 1 ,
  \label{eq:FinalStressTensor}
\end{eqnarray}
with the understanding that the abbreviation $P_{\rm bulk}$ means the
evaluation of the bulk equation of state \textit{for the local
  density} $\rho(\vec{r})$.

When adjusting $U_\textrm{s}({\bf r})$ to reproduce a desired contact
angle on a non-corrugated substrate, the canonical ensemble is more
convenient. The Euler-Lagrange equation is then derived by minimizing
the Helmholtz free energy, obtained from
Eq.~(\ref{eq:grandcanfree}). This must be done under the constraint
that the total number of molecules $N$ in the system remains fixed.
In this case, the relationship between the mean field felt by the
molecule and the local density is still described by
Eq.~(\ref{eq:SCF_GC1}). However the counterpart of
Eq.~(\ref{eq:SCF_GC2}) in the canonical ensemble is
\begin{equation}
\rho({\bf r}) = \frac{N \exp(- W({\bf r}))}
{\int d{{\bf r^\prime}}\exp(- W({\bf r^\prime}))} .
\label{eq:SCF_C}
\end{equation}

\subsection{Parameterization}
\label{ssec:param}

At ambient conditions ($T = 298.15 \, \mathrm{K}$, $P = P_{\rm a} = 0.1013 \, 
\mathrm{MPa}$, $P / k_B T = 0.0246 \, (\mathrm{nm})^{-3}$) liquid water 
has~\cite{Lide_2004} a density (number of molecules per volume) of $\rho_{\rm a} 
= 33.33 \, (\mathrm{nm})^{-3}$, and a bulk modulus $K = \rho \partial P / 
\partial \rho = 2.21 \, \mathrm{GPa}$, or $K / k_B T = 537 \, 
(\mathrm{nm})^{-3}$. From the equation of state, 
Eq.~(\ref{eq:pressure_in_bulk}), combined with the simple model free energy, 
Eq.~(\ref{eq:Helmbulk}), we find for pressure and modulus
\begin{eqnarray}
  \frac{P}{k_B T} & = & \rho + \frac{v}{2} \rho^2 + \frac{2}{3} w \rho^3 ,
  \label{eq:ParamPressureEquation} \\
  \frac{K}{k_B T} & = & \rho + v \rho^2 + 2 w \rho^3 .
  \label{eq:ParamModulusEquation}
\end{eqnarray}
Insertion of the the values for $\rho$, $P / (k_BT)$, and $K /
(k_BT)$, as given above for ambient conditions, gives rise to a set of
two linear equations for $v$ and $w$. Its solutions is $v \approx
-1.09 \, (\mathrm{nm})^3$, and $w \approx 0.023 \, (\mathrm{nm})^6$.
Throughout the study, we consider the temperature $T = 298.15 \,
\mathrm{K}$.

Bulk phase coexistence between liquid and vapor occurs at a
significantly lower pressure $P_{\rm c} < P_{\rm a}$. Denoting the
densities in the liquid and vapor phases with $\rho_{\rm l}$ and
$\rho_{\rm v}$, respectively, the conditions for phase coexistence are
(i) equality of the pressure, i.~e.
\begin{equation}
\rho_{\rm l} + \frac{v}{2}\rho_{\rm l}^2 + \frac{2}{3}w \rho_{\rm l}^3 
= \rho_{\rm v} + \frac{v}{2}\rho_{\rm v}^2 + \frac{2}{3}w \rho_{\rm v}^3 ,
\label{eq:pressure}
\end{equation}
and (ii) equality of the chemical potential, i.~e.
\begin{equation}
\rho_{\rm l} \exp \left( v \rho_{\rm l} + w \rho_{\rm l}^2 \right) =
\rho_{\rm v} \exp \left( v \rho_{\rm v} + w \rho_{\rm v}^2 \right)
\label{eq:ChemPotCoex}
\end{equation}
(cf. Eq.~(\ref{eq:chem_pot_def})). To determine $\rho_{\rm l}$ and
$\rho_{\rm v}$ one needs to solve this set of equations
numerically. For the parameters given above, this results in a liquid
density that is slightly decreased relative to the ambient value, but
essentially the same, due to the large value of $K$ (and identical
within the given accuracy). The vapor density is found to be
$\rho_{\rm v} \approx 8.7 \times 10^{-4} \, (\mathrm{nm})^{-3}$,
\rev{while the coexistence pressure (in units of $k_B T$) is found to
  be $P_{\rm c} / k_B T \approx 8.7 \times 10^{-4} \,
  (\mathrm{nm})^{-3}$, which is equivalent to $P_{\rm c} \approx
  3.6\, \mathrm{kPa}$.}

Since $K$ is very large, one can alternatively solve the problem with
fairly good accuracy in a simpler way, and this is the way how we
actually determined our parameters: We assume from the outset that at
coexistence the liquid density $\rho_{\rm l}$ takes the value
$\rho_{\rm a}$, i.~e. we neglect the tiny decrease in density that
results from decreasing the pressure from one atmosphere to the
coexistence value. We then consider
Eqs.~(\ref{eq:ParamModulusEquation}), (\ref{eq:pressure}), and
(\ref{eq:ChemPotCoex}) as three equations for the three unknowns $v$,
$w$, and $\rho_{\rm v}$, which are solved simultaneously, while the
coexistence pressure is determined after this from
Eq.~(\ref{eq:ParamPressureEquation}). \rev{This results in $v$, $w$,
  $\rho_{\rm v}$, and $P_{\rm c}$ values that are slightly shifted
  but, within the given accuracy, identical to the values already
  given. The precise values of the parameters at which the study was
  performed are $v = -1.086641506142464 \, (\mathrm{nm})^3$ and $w =
  0.023102120829070 \, (\mathrm{nm})^6$. The vapor density and the
  coexistence pressure are reasonably close to their experimental
  counterparts~\cite{Lide_2004}, $\rho_{\rm v} \approx 7.7 \times
  10^{-4} \, (\mathrm{nm})^{-3}$ and $P_{\rm c} \approx 3.2\,
  \mathrm{kPa}$. It should be emphasized that, taking into account the
  simplicity of the free-energy functional, it can be hardly expected
  to reproduce real experimental data more faithfully.  Nevertheless,
  the most important properties of bulk water have been taken into
  account.}

To estimate the parameter $\kappa$ we consider a flat liquid-vapor
interface in the absence of a substrate potential. The one-dimensional
density profile $\rho(z)$ then varies between $\rho_{\rm v}$ for $z
\to - \infty$ and $\rho_{\rm l}$ for $z \to + \infty$. The surface
tension $\gamma$ is then given by~\cite{JonesBook, Yang1976}
\begin{equation}
  \frac{\gamma}{k_BT} = \kappa \int_{-\infty}^{+\infty} dz 
  \left(\frac{d\rho(z)}{dz}\right)^2.
  \label{eq:surface_tension}
\end{equation}

To proceed, we note that the problem of finding the one-dimensional density 
profile $\rho(z)$ is mathematically identical to solving the one-dimensional 
equation of motion of a particle in an external potential, by identifying 
$\Omega$ with the action integral. The standard method to solve such a 
problem~\cite{LL_EL_76} is to write down the equation for energy conservation 
and to separate variables. This allows us, after eliminating $\mu$, and some 
straightforward algebra, to transform the integral to the form
\begin{equation}
  \frac{\gamma}{k_B T} = \sqrt{\kappa} \int_{\rho_{\rm v}}^{\rho_{\rm l}} d\rho \;
  \sqrt{2 \phi (\rho) }
  \label{eq:GammaIntegral}
\end{equation}
with
\begin{eqnarray}
  \phi (\rho) & = & \rho \ln \left( \frac{\rho}{\rho_{\rm l}} \right)
  + \left( \rho_{\rm l} - \rho \right)
  + \frac{v}{2} \left( \rho_{\rm l}^2 + \rho^2 \right)
  \nonumber
  \\
  & &
  + \frac{w}{3} \left( 2 \rho_{\rm l}^3 + \rho^3 \right)
  - \rho \left( v \rho_{\rm l} + w \rho_{\rm l}^2 \right) .
  \label{eq:GammaEvaluated}
\end{eqnarray}

Numerical evaluation yields 
\begin{equation}
\frac{\gamma}{k_B T} = \sqrt{\kappa} \cdot 199.26 \, \textrm{nm}^{-9/2}.
\end{equation}

Experimentally~\cite{Kayser_1976} $\gamma = 0.07275\, \mathrm{N/m}$ or 
$\gamma/k_B T = 17.673 \, \textrm{nm}^{-2}$ and we thus obtain $\kappa = 
0.007866 \, (\mathrm{nm})^5$.

The interfacial width $h$ can be estimated approximately by assuming a
profile that is linear on a scale $h$ and flat otherwise. This yields
(see Eq.~\ref{eq:surface_tension})
\begin{equation}
  \frac{\gamma}{k_B T} \approx \frac{\kappa}{h} (\rl - \rv)^2
  \label{eq:GammaScaling}
\end{equation}
or $h \approx 0.5 \, \textrm{nm}$. This would however impose the need
of a very fine discretization grid. We therefore increase $\kappa$ to
the value $\kappa = 0.031478342 \, (\mathrm{nm})^5$ (actual value used
in the calculations). This implies $\gamma = 0.1455 \, \textrm{N/m}$
and $h \approx 1 \, \textrm{nm}$ (since $\gamma \propto \sqrt{\kappa}$
it follows from Eq.~\ref{eq:GammaScaling} that $h \propto
\sqrt{\kappa}$). This allows us to perform the calculations employing
a lattice spacing of $0.1 \, \textrm{nm}$, which reduces the
computational effort significantly, compared to the otherwise needed
value of $0.05 \, \textrm{nm}$. Furthermore, according to the
simulation results of Ref.~\cite{Netz}, an interfacial width of $1 \,
\textrm{nm}$ is also expected to be physically more realistic.

\subsection{Corrugated substrates}
\label{ssec:subs}

We model the corrugation geometry by a function $\rho_0 (\vec{r})$,
the number density of the substrate atoms. We assume that this
function is constant (i.~e.  $\rho_0 (\vec{r}) = \rho_0$) within the
space occupied by the substrate and zero elsewhere. Describing the
substrate in terms of continuum theory (just as the water), and
assuming pairwise additive interactions, we can hence write
\begin{equation}
U_{\rm s} (\vec{r}) = \int d{\bf r^\prime} \; U_{\rm sw}
\left( \left\vert \vec{r} - \vec{r^\prime} \right\vert \right)
\rho_0 \left( \vec{r^\prime} \right) ,
\label{eq:SubstrateConvolution}
\end{equation}
where $U_{\rm sw}$ is the interaction potential between a substrate
atom and a water molecule (in units of $k_B T$). For the latter, we
use a 12-6 Lennard-Jones (LJ) potential:
\begin{equation}
U_{\rm sw} (r) = \left\{
\begin{array}{l l}
\frac{4 \epsilon}{k_B T}
\left[ \left( \frac{\sigma}{r} \right)^{12} -
\left( \frac{\sigma}{r} \right)^{6} \right] 
& r \le r_{\rm c} , \\
0 & r > r_{\rm c} ,
\end{array}
\right.
\end{equation}
where $\epsilon$ and $\sigma$ are the characteristic energy and length
scale, respectively. In our calculations we use $\sigma = 0.3 \,
\mathrm{nm}$ and $r_\textrm{c} = 5 \, \mathrm{nm}$, while the
potential depth $\epsilon$ is adjusted to reproduce the desired
contact angle on a flat substrate. More precisely, it is only the
combined parameter $\tilde \epsilon = \epsilon \rho_0 / k_B T$ that
matters, and this is what we vary.

\begin{figure}
  \centering
  \includegraphics[width=8.5cm]{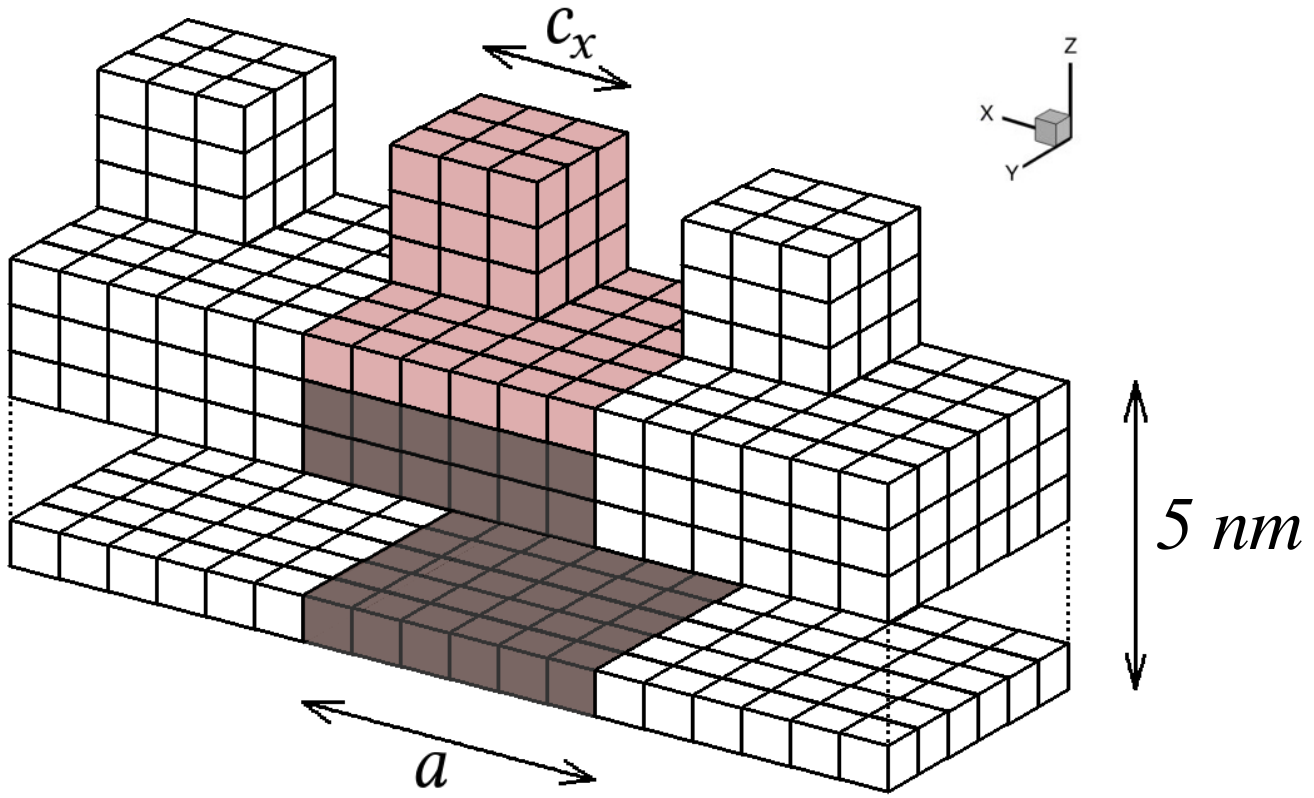}
  \caption{(color online) Set-up for defining the substrate
    potential. The substrate explicitly present in the calculation box
    is shown in pink. To construct the substrate potential a
    "bulk'' of the substrate is taken into account (brown). The
    obtained substrate unit is subsequently replicated (for clarity
    only two images in $x$-direction are shown in white) based on the
    cut-off $r_\textrm{c}$ of the LJ potential.}
  \label{fig:sub_create}
\end{figure}

As the substrate structure is strictly periodic in the $x$ and $y$
directions (parallel to the surface), it is sufficient to restrict
attention to one unit cell of the corrugation. This results in a
simulation cell with periodic boundary conditions in $x$ and $y$
directions. The space $z \le 0$ corresponds to the ``bulk'' substrate,
i.~e. $\rho_0 (\vec{r}) = \rho_0$ whenever $z < 0$. In contrast, the
space $z > 0$ corresponds to the corrugated part, i.~e. here we have
$\rho_0 (\vec{r}) = 0$ for the space left for the water, while $\rho_0
(\vec{r}) = \rho_0$ in the regions where there is substrate
material. In $z$ direction we hence choose a box size that is large
enough to faithfully model (i) the interactions with all the substrate
atoms, and (ii) the thermal equilibrium with a bulk reservoir of
water. This requires an extension of at least $r_{\rm c}$ in the
negative $z$ direction, and an extension in positive $z$ direction
that is significantly larger than the size of the corrugation, plus
the interfacial width. A typical unit cell is illustrated in
Fig.~\ref{fig:sub_create} in pink (one repeat unit of corrugation) and
brown (``bulk'' of the substrate).

The continuum theory is discretized in terms of a simple cubic grid
with lattice spacing $\Delta L = 0.1 \, \mathrm{nm}$, and for
simplicity we choose the same lattice structure and constant in both
the space occupied by water and substrate. The integral in
Eq.~(\ref{eq:SubstrateConvolution}) is then discretized by a sum,
which has to be evaluated once and for all at the beginning of the
calculation (i.~e. with negligible computational cost). This formula
is however used only for those sites $\vec{r}$ that are not occupied
by the substrate.

In order to avoid complicated boundary conditions at the substrate
surface, we allow (in principle) the water to penetrate into the
substrate. In other words, the equations are solved everywhere in the
computational domain, including the space occupied by the substrate.
In this latter part, however, the water density is very small, since
on the substrate sites we introduce a very large potential that
strongly penalizes water penetration:
\begin{equation}
U_\textrm{s}(\vec{r})|_{\rho_0 (\vec{r}) > 0} = U_\textrm{cap} ,
\label{eq:surf_cap}
\end{equation}
where $U_\textrm{cap}$ is set to the maximum of all the $U_\textrm{s}$
values on the sites not occupied by the substrate. \rev{This potential
  replaces Eq.~(\ref{eq:SubstrateConvolution}) on all sites $\vec{r}$
  within the substrate; note that in this case
  Eq.~(\ref{eq:SubstrateConvolution}) would yield an infinite
  potential value and thus an ill-defined model.}

The space occupied by the "bulk'' of the substrate and the replicas of
the unit cell (brown and white in Fig.~\ref{fig:sub_create},
respectively) is excluded from the iteration scheme. The solution is
only performed within the calculation box encapsulating the
corrugation unit (pink in Fig.~\ref{fig:sub_create}) and the space
above it. The boundary conditions will be discussed in
Sec.~\ref{ssec:numscheme}.

The above description of the substrate is acceptable for a generic
study, since the interactions with water are in any case parameterized
in a top-down fashion. It should be noted that the integral in
Eq.~(\ref{eq:SubstrateConvolution}) can be done analytically only for
fairly simple geometries~\cite{GB_ER_2008, TB_LJ_2008}, and hence we
do it numerically as outlined above. This approach allows us to
implement essentially any desired substrate geometry. When considering
real substrates $U_\textrm{s}({\bf r})$ must be defined through more
elaborate summation schemes~\cite{Steele2, Doros} to avoid, e.~g.,
reduction of the adhesion strength due to the cut-off.  In this
context, it should be noted that one may alternatively view our
procedure to calculate $U_{\rm s}$ as a way to sum up the potential
contributions from individual atoms located on a simple-cubic lattice
of spacing $0.1 \textrm{nm}$. Therefore, the fact that the
discretization of the integral Eq. \ref{eq:SubstrateConvolution}
introduces a slight corrugation of scale $0.1 \textrm{nm}$ should not
necessarily be viewed as an undesirable artifact. Rather, this can be
interpreted as a means to take effects of atomic-scale corrugation,
like, e.~g., interface pinning, into account --- of course only in a
very simple and certainly not quantitatively reliable way. 

The dimensions of the calculation cell ($L_x$, $L_y$ and $L_z$) depend
on the type of corrugation. For striped substrates with stripes
oriented in the $y$ direction, we can exploit translational invariance
and set $L_y$ to a very small value ($L_y = 1.6 \, \mathrm{nm}$). In
this case $L_x$ is given by the periodicity of the stripes $a$,
i.~e. $L_x = a$. The width of the stripe is controlled by the
parameter $c_x$ (cf. Fig.~\ref{fig:sub_create}). In the case of
pillared substrates, the $x$ and $y$ dimensions of the box are equal,
and identical to the period of the pillars $a$, i.~e. $L_x = L_y =
a$. The pillars have a quadratic cross section of size $c_x \times
c_y$ with $c_x = c_y$. For both substrates, $c_x$ and $a$ are the
control parameters to vary the substrate geometry. The height of the
corrugations and the dimension of the box in $z-$direction are set to
$2 \, \mathrm{nm}$ and $L_z = 16.4 \, \mathrm{nm}$, respectively.

\subsection{Numerical scheme}
\label{ssec:numscheme}

To solve the set of Eqs.~(\ref{eq:SCF_GC1}) and~(\ref{eq:SCF_GC2}) (or
Eqs.~(\ref{eq:SCF_GC1}) and~(\ref{eq:SCF_C})) a real-space method is
employed, akin to numerical schemes developed for the treatment of the
SCF formalism in polymers~\cite{Drolet, MarcusFriederike,
  FredricksonReview}. The system is discretized through a regular
cubic grid, with lattice spacing $\Delta L = 0.1 \, \mathrm{nm}$, and the
set of equations is solved to obtain the values of $\rho({\bf r})$ and
$W({\bf r})$ on the nodes of this lattice. The non-local term $\kappa
\vec{\nabla}^2 \rho({\bf r})$ is approximated by a finite-difference
method. The employed central-difference stencil has $3$ points in each
of the three dimensions~\cite{AndersonAero}.

To calculate finite differences at the boundaries of the calculation
cell, we utilize Dirichlet conditions in the $z$ direction, and
periodic boundary conditions in the $x$ and $y$ directions. For the
former, we introduce two layers located at $z = 0$ and at $z = L_z$
(the ``bottom'' and the ``top'' layer), at which we prescribe the
values of $\rho$. The nodes inside the cell are then located at $z =
\Delta L / 2, 3 \Delta L / 2, \ldots, L_z - \Delta L / 2$, and the
equations are only solved at these inner nodes. At the bottom we set
the water density to zero, $\rho_\textrm{bot} = 0$, which corresponds
to an infinitely repulsive surface potential acting on that layer. In
the case of grand-canonical calculations, the density at the top layer
is set to $\rho_\textrm{top} = 33.40 \, (\mathrm{nm})^{-3}$. We 
deliberately choose this value (which is larger than the coexistence 
density) to impose on the top of the system a moderate pressure, $\Pmax
 \simeq 50 \, \mathrm{atm}$ (see Sec.~\ref{ssec:Corrugated} for more details). 
For calculations in the canonical ensemble, where the droplet fits completely 
into the cell, we rather set $\rho_\textrm{top} = \rho_\textrm{v}$, i.~e. 
the vapor density.

To calculate three-dimensional integrals (e.~g. free energy, partition
function, etc.) we use Simpson's integration method with semi-open
($x$ and $y$ directions) or open ($z$ direction)
boundaries~\cite{WHP_SAT_WTV_2007}.

The numerical solution of the set of Eqs.~(\ref{eq:SCF_GC1})
and~(\ref{eq:SCF_GC2}) (or (\ref{eq:SCF_C})) proceeds via simple
iteration. Starting from a field $W (\vec{r})$, a new density profile
is calculated by inserting $W$ into the right-hand side of
Eq.~(\ref{eq:SCF_GC2}) (or (\ref{eq:SCF_C})). From this new profile, a
new field $W_{\rm new} (\vec{r})$ is obtained via insertion into the
right-hand side of Eq.~(\ref{eq:SCF_GC1}).The field $W$ is then
updated not by simply replacing it with $W_{\rm new}$, but rather with
the linear combination $\lambda W_{\rm new} + (1 - \lambda) W$. Here
$\lambda$ ($0 < \lambda \le 1$) is a relaxation parameter introduced
to avoid numerical instabilities in the iteration. For our system, we
found that a fairly small value ($\lambda = 10^{-3} \ldots 2 \times
10^{-3}$) is needed. After a few hundred thousand steps the iteration
has converged, meaning that the relative change in the field, summed
over all sites, is smaller than $10^{-25}$. The program that we
developed for this purpose is parallelized via the Message Passing
Interface (MPI) standard, and publicly
available~\cite{Tretyakov_MF_github}.

\section{Wetting at flat and corrugated substrates}
\label{sec:water-vapor}
%

\subsection{Flat substrate}
\label{ssec:canonical}

As a first step, we need to establish a connection between the
interaction parameter $\tilde \epsilon$ and the contact angle $\tflat$
of droplets on a flat substrate (i.~e. Young's angle). For this
purpose we perform a series of calculations in the canonical ensemble,
and study droplets of different size (or different number of water
molecules $N$) at constant $\tilde \epsilon$. For any finite size of
the droplet, the contact angle is affected by line tension
contributions~\cite{KB_BB_SKD_2011}, and hence the true asymptotic
contact angle is obtained only after extrapolation to droplets of
infinite size. Since the line tension effects are different for
spherical and cylindrical droplets, it is advisable to study both
types, such that the extrapolation can be done in a more reliable
way. The spherical droplets are studied in a cell with quadratic cross
section, $L_x = L_y$. The cylindrical droplets are aligned in $y$
direction and infinitely long, such that we can make use of
translational invariance to set $L_y$ to the small value
$1.6 \, \mathrm{nm}$. We study the two values $\tilde \epsilon = 150
 \, (\mathrm{nm})^{-3}$ and $100 \, (\mathrm{nm})^{-3}$. In order to check that 
the solutions are
well-converged, we start the iteration from two very different
starting configurations (droplets with contact angles $90^\circ$ and
$180^\circ$) and confirm that the final results are identical.

\begin{figure}
 \centering
  \includegraphics[width=8.5cm]{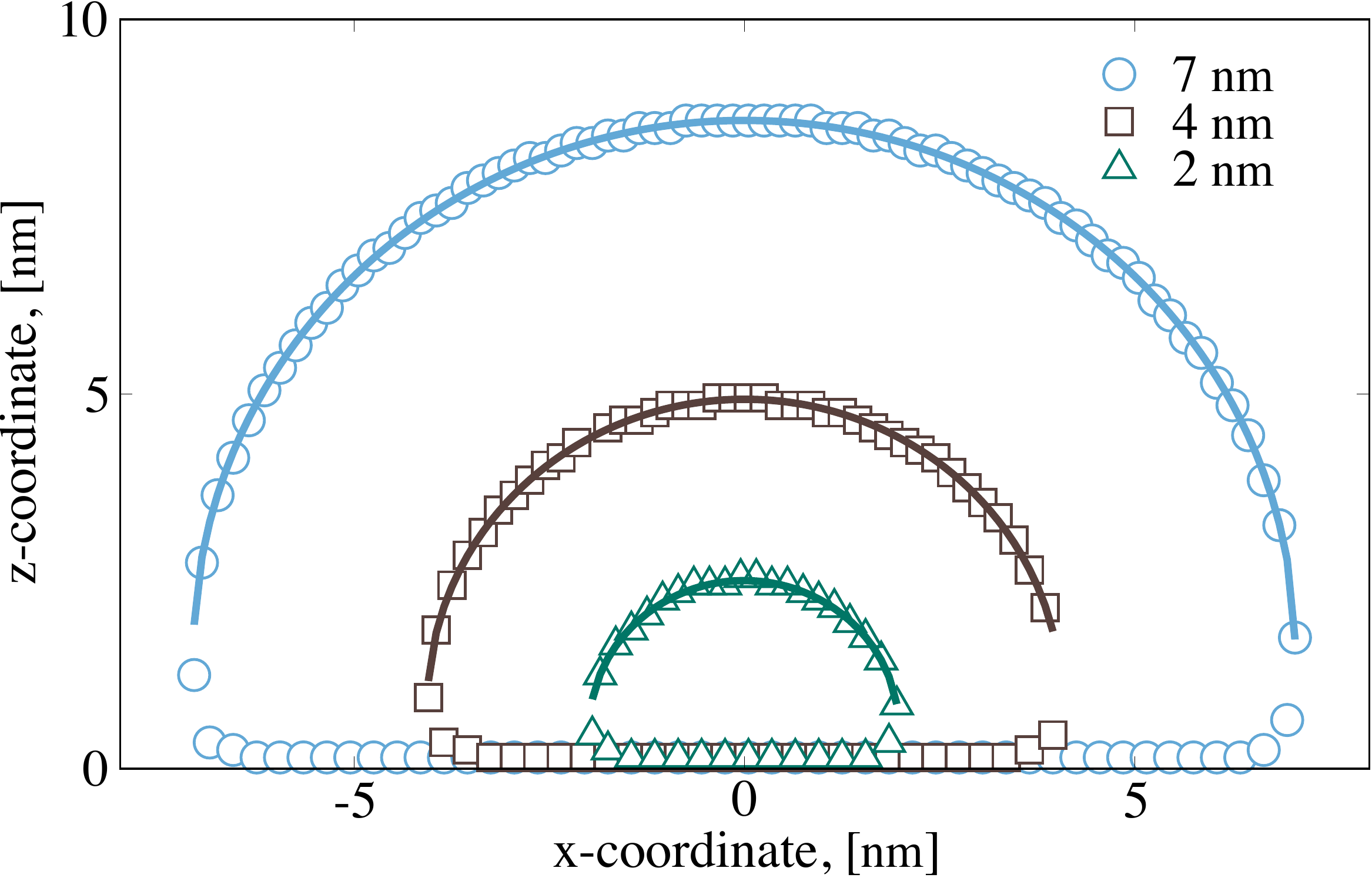}
 \caption{Shapes of droplets with radii of ca. $2$, $4$
   and $7\mathrm{nm}$ (symbols) and the corresponding fits of their upper
   parts with a spherical cap (solid lines). The contact angle
   $\tflat$ then results from the fit parameters.}
 \label{fig:ca_calc}
\end{figure}

\begin{figure}
 \centering
  \includegraphics[width=8.5cm]{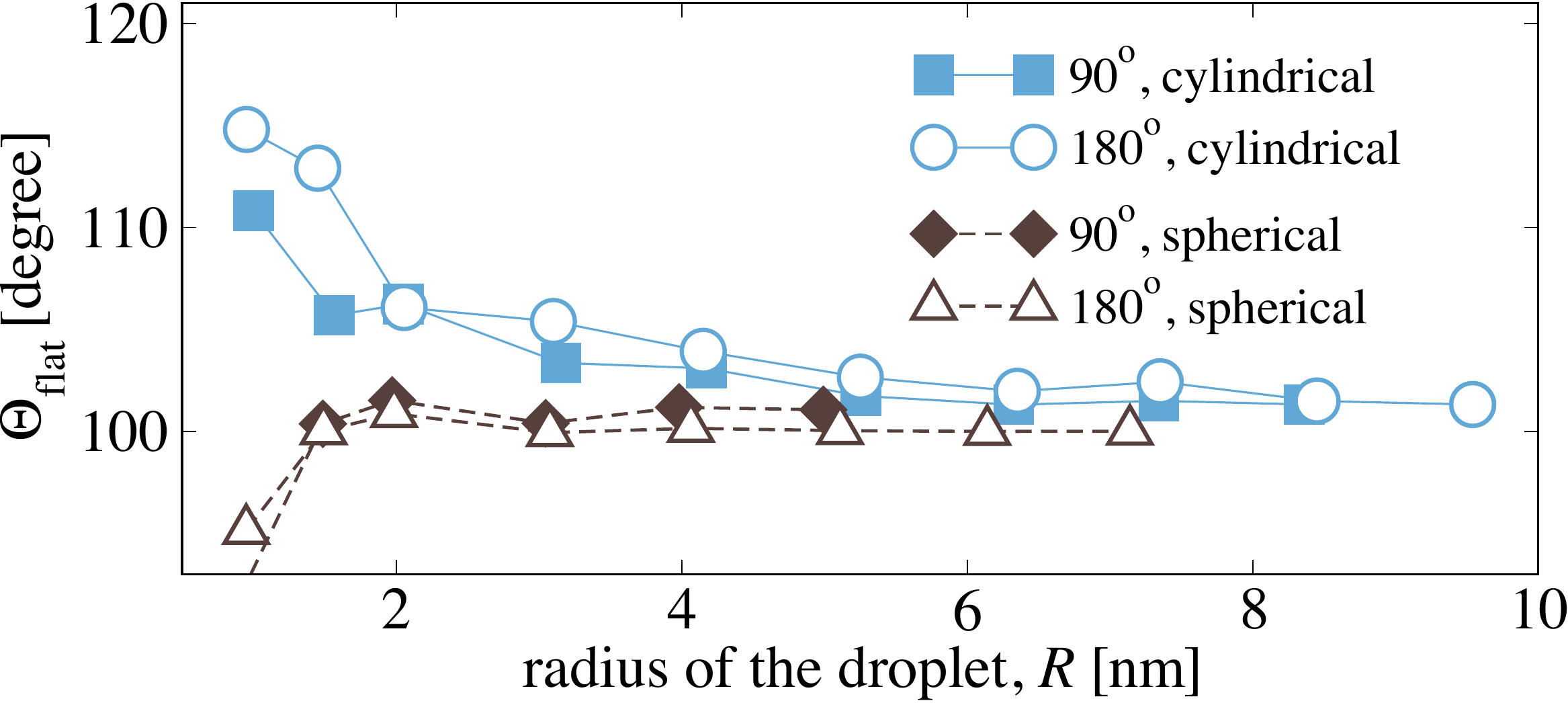}
 \caption{Comparison of the contact angle of cylindrical
   (two-dimensional) and spherical (3D) droplets on a flat substrate of 
   $\tilde \epsilon = 150 \, (\mathrm{nm})^{-3}$ as a function of the 
   droplet's radius, $R$. Initial configurations with contact
   angle of $180^\circ$ (full cylinder and sphere) and $90^\circ$
   (half-cylinder and semi-sphere) are utilized.}
 \label{fig:ca_comp}
\end{figure}

After relaxation to the equilibrium shape, we measure the contact
angle. To this end, we first obtain a two-dimensional $xz$-density map
in a plane through the droplet's center of mass. Then, the
liquid-vapor interface is localized according to the criterion $\rho =
(\rl + \rv)/2$. As an example, Fig.~\ref{fig:ca_calc} presents the
resulting shapes of drops of various sizes. To avoid uncertainties
near the substrate, we apply a spherical-cap approximation to the
upper part of the shapes (solid lines in Fig.~\ref{fig:ca_calc}). We
thus obtain fitted radii and fitted positions of the circles' centers,
from which the contact angles can be directly inferred. 

As an example of the extrapolation to infinite droplet size, we
present in Fig.~\ref{fig:ca_comp} the $\tflat$ data for $\tilde
\epsilon = 150 \, (\mathrm{nm})^{-3}$, for both cylindrical and
spherical drops, and the two initial configurations that we
studied. For the other amplitudes $\tilde \epsilon$ the behavior is
similar. As the contact angle converges from above for cylinders, and
from below for spheres, the extrapolated value can be obtained fairly
accurately. Data for radii $\le 3 \, \mathrm{nm}$ should be discarded,
since in this regime the fitting procedure becomes rather
unreliable. This is hardly surprising, in view of the interfacial
thickness of $1 \, \mathrm{nm}$. Conversely, for radii $\ge 6 \,
\mathrm{nm}$ the asymptotic behavior is essentially reached. Averaging
over these large droplets, we obtain $\tflat = 101.26^\circ$ and
$120.03^\circ$ for $\tilde \epsilon = 150 \, (\mathrm{nm})^{-3}$ and
$100 \, (\mathrm{nm})^{-3}$, respectively. In the following, we refer
to these values as $100^\circ$ and $120^\circ$ for simplicity. Our
study therefore spans a region from weakly to moderately strongly
hydrophobic substrate materials.

\subsection{Corrugated substrates}
\label{ssec:Corrugated}

\begin{figure}
 \centering
  \includegraphics[width=8.5cm]{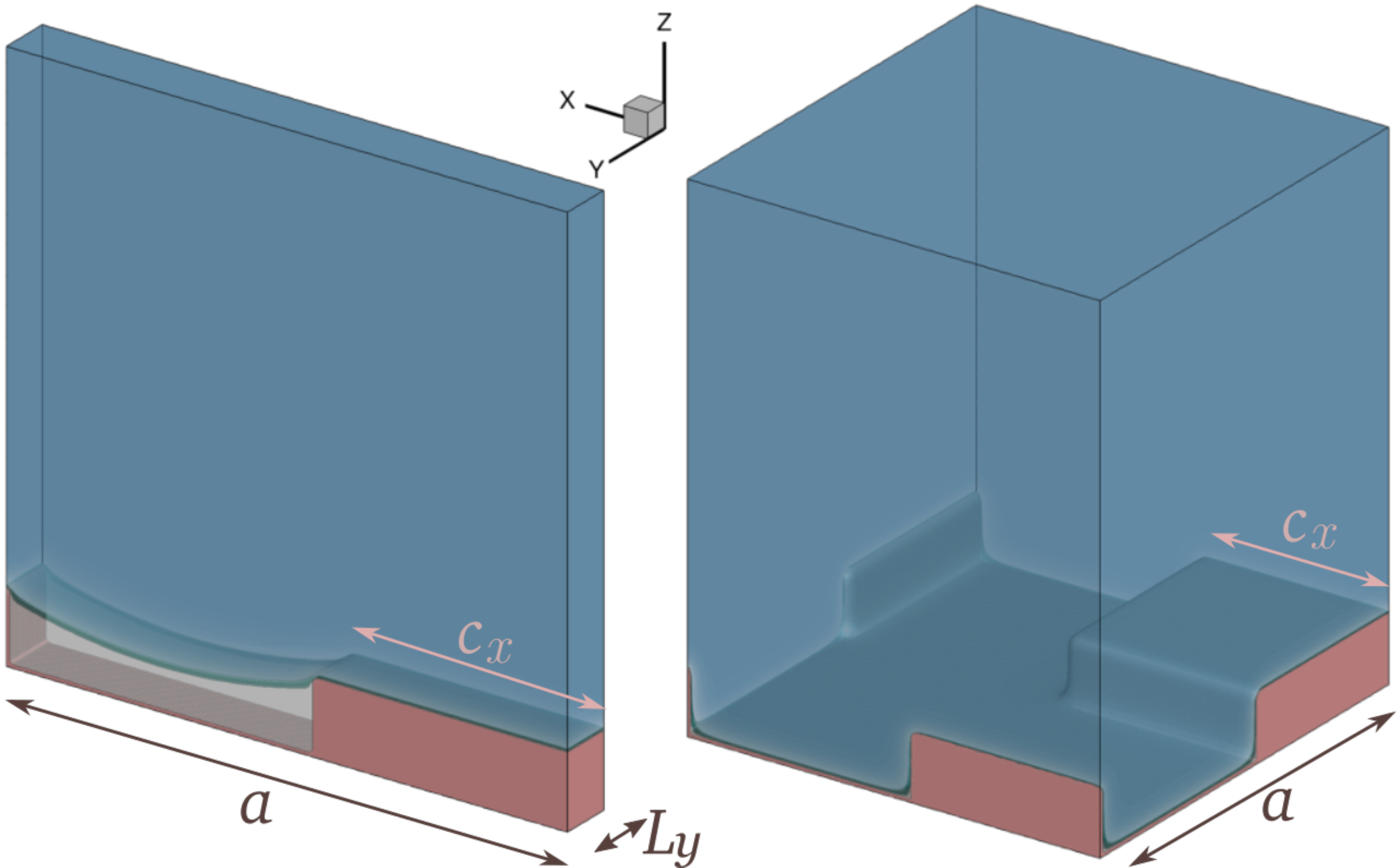}
 \caption{(color online) The parameters of the substrate of striped (left) and
   pillared (right) corrugation type. The converged snapshots
   represent Cassie (left) and Wenzel (right) states. Parts of the
   corrugations re-enter the calculation box through periodic boundary
   conditions.}
 \label{fig:geom}
\end{figure}

For striped surfaces, the fraction $f$ of surface that is covered with
liquid in the Cassie state is obviously given by $c_x /
a$. Furthermore, the contour length of the three-phase line is given
by $L_{\rm c} = 2 L_y$, while the total area $A$ is $A = a L_y$ (see
Fig.~\ref{fig:geom}, left). The force-balance equation,
Eq.~(\ref{eq:hj_gen}), can hence be re-written as
\begin{equation}
  \frac{c_x}{a} = \frac{b}{a} \cos \Theta_\textrm{max} + 1; \;\; \;\;
  b = \frac{2 \gamma}{\Pmax} .
  \label{eq:hj_stripes}
\end{equation}

\begin{figure*}[t]
  \subfloat[]{\label{fig:eps015_gr}\includegraphics[width=8.5cm]%
    {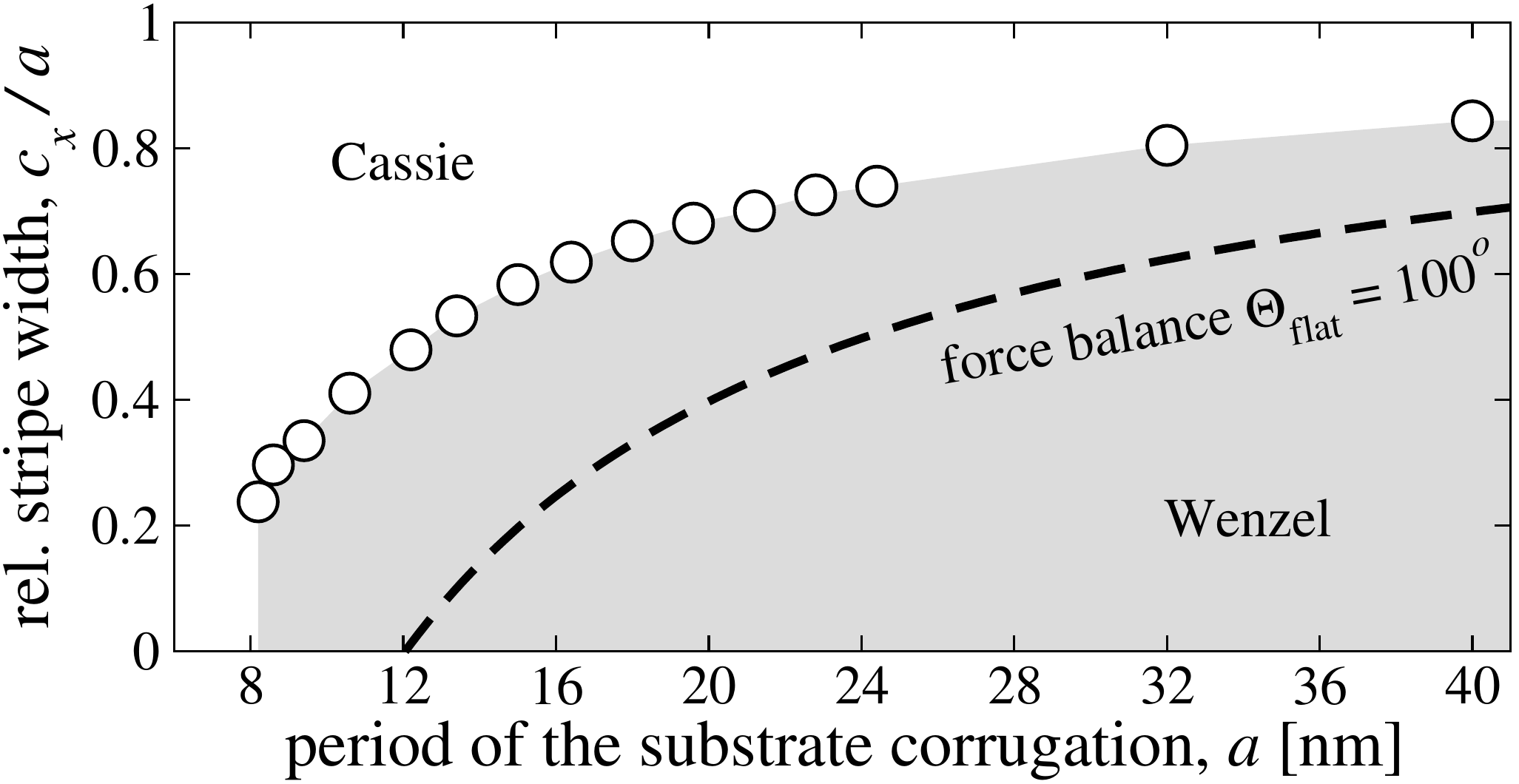}}\hspace*{0.3cm}
  \subfloat[]{\label{fig:eps015_pi}\includegraphics[width=8.5cm]%
    {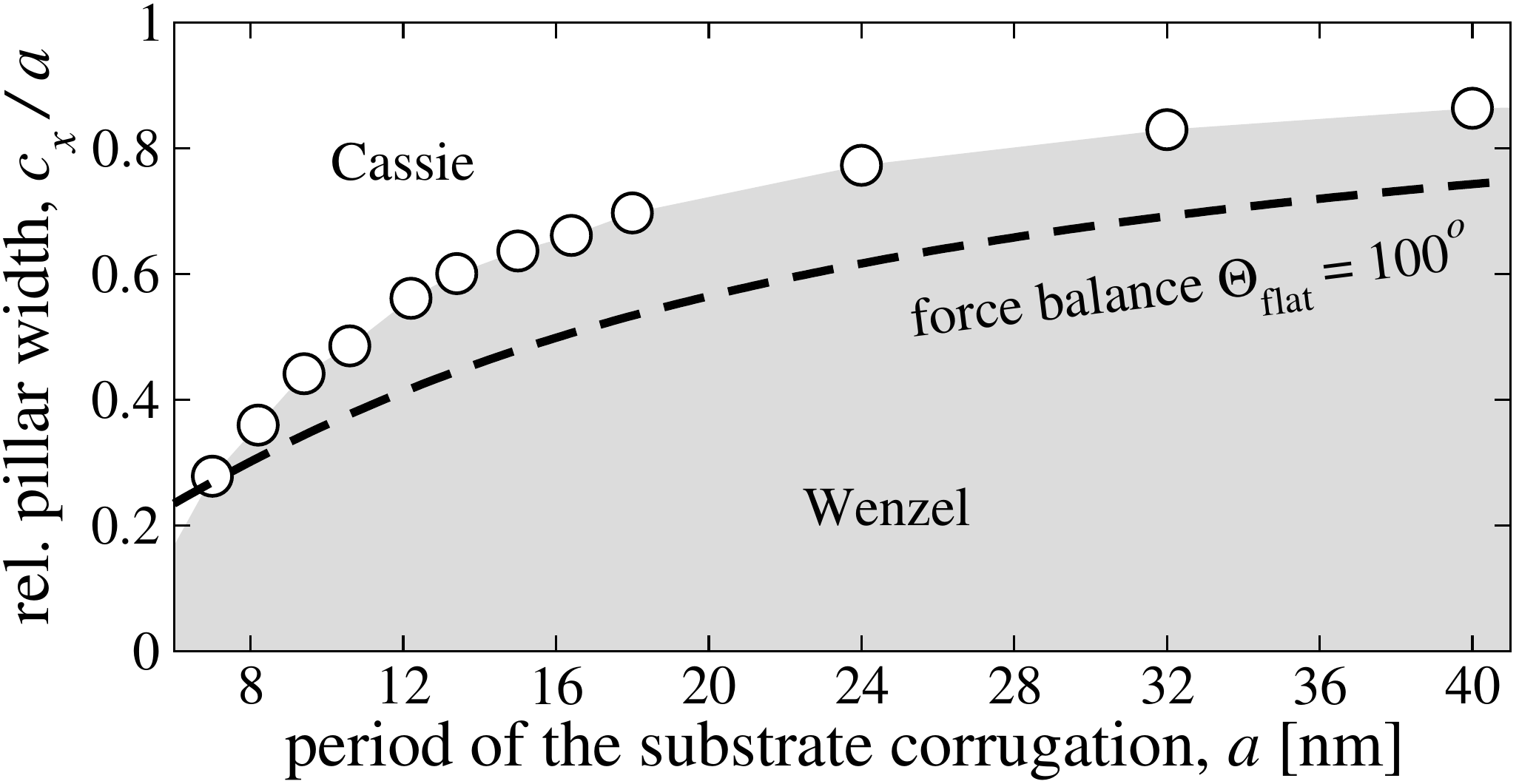}}
  
  \subfloat[]{\label{fig:eps010_gr}\includegraphics[width=8.5cm]%
    {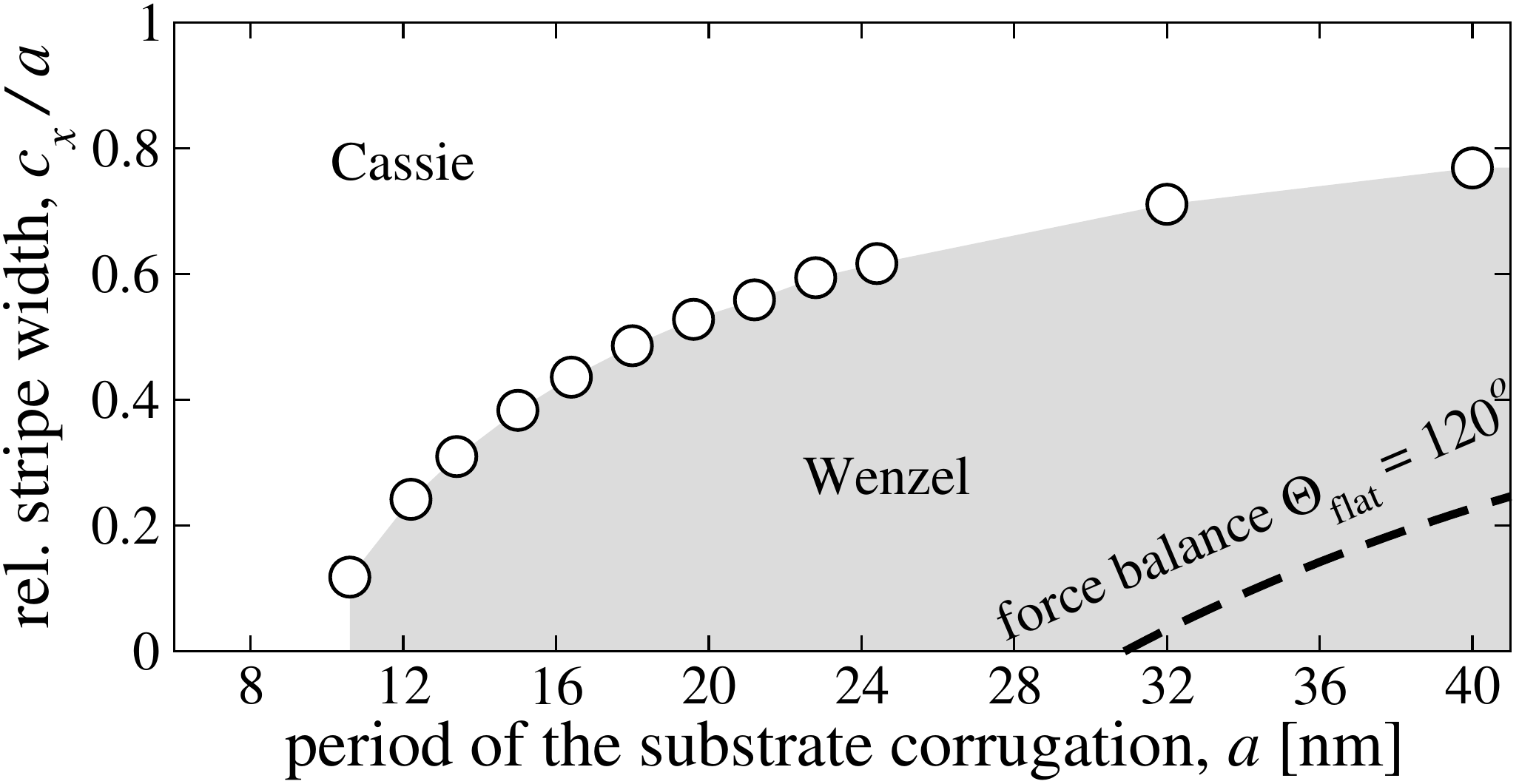}}\hspace*{0.3cm}
  \subfloat[]{\label{fig:eps010_pi}\includegraphics[width=8.5cm]%
    {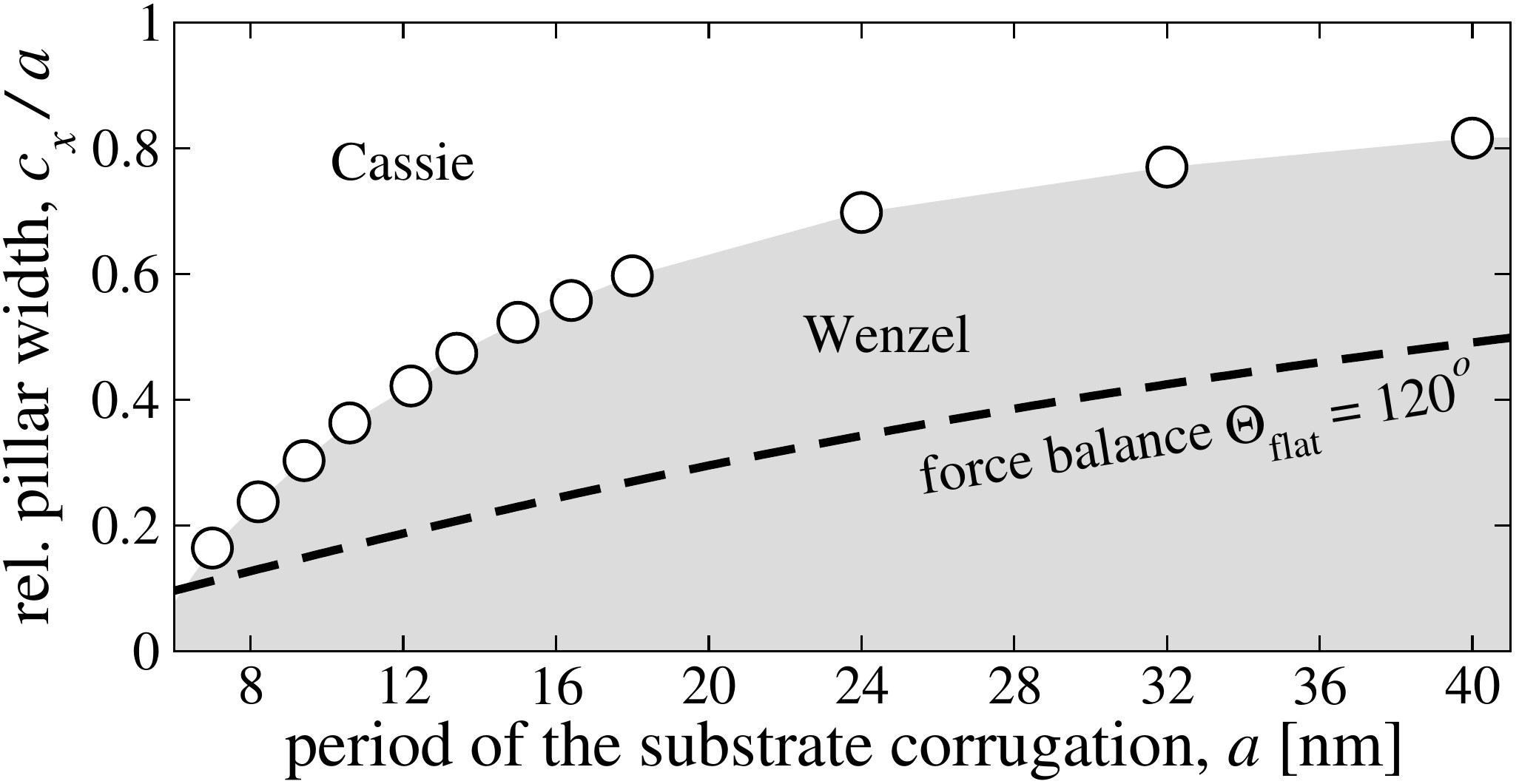}}

  \caption{State diagrams presenting the boundary between Cassie and
    Wenzel states, for striped (panels (a) and (c)) and pillared
    (panels (b) and (d)) substrates. The interaction strengths are
    $\tilde \epsilon = 150 \, (\mathrm{nm})^{-3}$ (panels (a) and
    (b)), and $\tilde \epsilon = 100 \, (\mathrm{nm})^{-3}$ (panels
    (c) and (d)). Open circles mark the limit of metastability of the
    Cassie state as calculated from DFT. The shaded regions in the
    lower right corners of the diagrams correspond to absolute
    stability of the Wenzel state. In the remaining white regions the
    Cassie state is either stable or metastable. Black dashed lines
    mark the corresponding limit of metastability as calculated from
    the macroscopic force-balance relation, employing the respective
    Young contact angle on a flat substrate of $\tflat = 100^\circ$
    (panels (a) and (b)), and $\tflat = 120^\circ$ (panels (c) and
    (d)). In all cases the hydrostatic pressure is $\Pmax \simeq 50 \,
    \mathrm{atm}$.  }
\label{fig:state_diagrams}
\end{figure*}

Similarly, for pillars (Fig.~\ref{fig:geom} right) we have $f = c_x^2
/ a^2$, $L_{\rm c} = 4 c_x$, $A = a^2$ and hence in this case
Eq.~(\ref{eq:hj_gen}) is re-written as
\begin{equation}
  \frac{c_x}{a} = \frac{b}{a} \cos \Theta_\textrm{max} + 
  \sqrt{\left( \frac{b}{a} \cos \Theta_\textrm{max} \right)^2 + 1}
  \label{eq:hj_pillars}
\end{equation}
(meaning of $b$ as above).

In order to test these relations, we pursue the following strategy: We
impose the boundary condition $\rho_\textrm{top} = 33.40 \,
(\mathrm{nm})^{-3}$ at the top of the calculation box
(cf. Sec.~\ref{ssec:numscheme}). We then evaluate the stress tensor,
Eq.~(\ref{eq:FinalStressTensor}), at the top layer. Since the profile
is rather flat at the top layer, the contribution $\propto \kappa$ can
safely be neglected. Hence
\begin{equation}
  \left. \frac{\tensor \Pi}{k_B T} \right\vert_{\rm top}
  = P_{\rm bulk} (\rho_{\rm top}) \tensor 1 ,
\end{equation}
and therefore we effectively impose a boundary condition of constant
pressure at the top surface, at a value of $P \simeq 50 \,
\mathrm{atm}$. \rev{This is much larger than the gas pressure at the
  bottom, which is of order $10^{-2} \mathrm{atm}$. For this reason,
  we can safely identify the pressure at the top surface with the
  excess pressure that enters Eq.~(\ref{eq:hj_gen}).} We then
systematically vary $c_x$ at constant $a$, starting with a fairly
large value of $c_x$. In this regime, the Cassie state is stable. Upon
decreasing $c_x$, a critical value is reached at which the Cassie
state collapses and rather the Wenzel state is observed. Obviously,
our calculations permit us to directly monitor this process. Our
iteration scheme starts with an interface located $\sim 1 \,
\mathrm{nm}$ above the top of the asperities; therefore the procedure
is expected to always converge to the (possibly metastable) Cassie
state, unless it is absolutely unstable. It should be recalled that
the purpose of the investigation is to find this limit of
metastability. Finally, this procedure is repeated for various values
of $a$, resulting in a state diagram in the plane $a$ vs. $c_x/a$. If
the phenomenological force balance holds, then the transition line
must be given by Eq.~(\ref{eq:hj_stripes}) for stripes, and by
Eq.~(\ref{eq:hj_pillars}) for pillars.

At this point, it should be noted that the parameters that enter the
force-balance relation are (i) the geometry data (input data of the
calculations), (ii) the pressure and the interfacial tension (known
accurately, see also end of Sec.~\ref{ssec:param}), and (iii) the
contact angle $\Theta_\textrm{max}$, which is not known. Due to the
diffuse nature of the interface and the nanoscopic corrugation
dimensions it is practically impossible to obtain a contact angle
directly from the simulation data. We therefore pursue the following
two strategies to do the comparison: (i) On the one hand, we simply
assume that the relevant contact angle is just the Young value
$\tflat$ as follows from macroscopic considerations
(cf. Appendix~\ref{sec:appendix}), and check how well the predicted
transition line matches the DFT result. This strategy relies on the
macroscopic assumption that contact angle hysteresis is of no
importance for ideal substrates, which has been demonstrated to be
valid at least for some systems~\cite{KK_MB_BML_2007,SH_MB_RS_2008}.
(ii) On the other hand, we solve the force-balance relation,
Eq.~\ref{eq:hj_gen}, for $\Theta_\textrm{max}$. We can hence determine
the contact angle as an effective parameter, to be evaluated along the
DFT phase transition line.

\begin{figure}
 \centering
  \includegraphics[width=8.5cm]{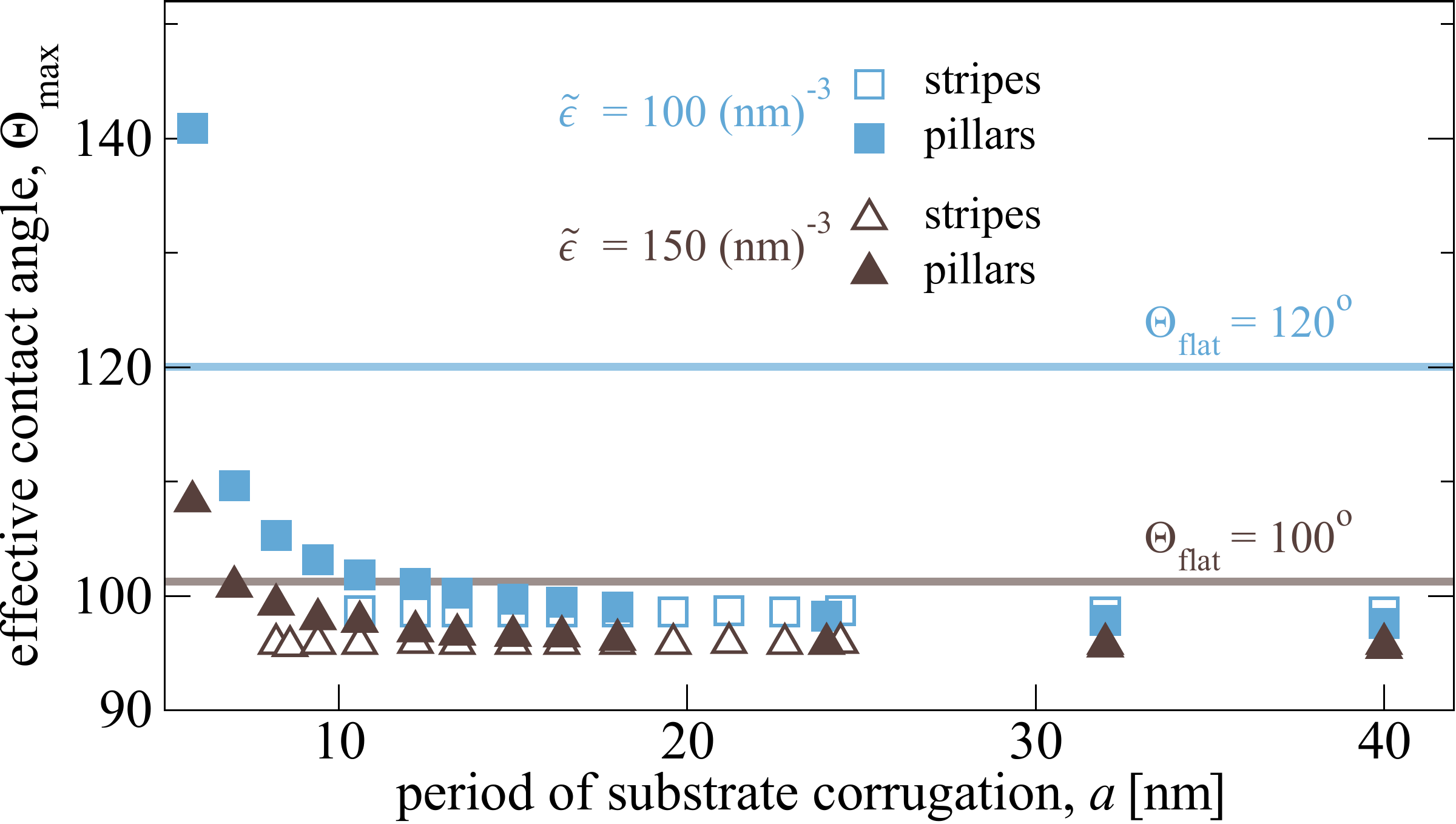}
  \caption{(color online) The effective contact angle
    $\Theta_\textrm{max}$ for the force-balance relation as a function
    of the period of the substrate corrugation $a$, which we use to
    parameterize the path along the Cassie-Wenzel transition line.}
\label{fig:ca_vs_a}
\end{figure}

The results of this analysis are summarized in
Figs.~\ref{fig:state_diagrams} and~\ref{fig:ca_vs_a}. The state
diagrams in Fig.~\ref{fig:state_diagrams} clearly show that the
force-balance relation gives a qualitatively reasonable description of
the Cassie-Wenzel transition even on the nanoscale. However, the
attempt to describe the phenomenon quantitatively by just employing
the Young contact angle (black dashed lines) fails. The deviations
increase systematically with increasing hydrophobicity, as clearly
shown in Fig.~\ref{fig:ca_vs_a}, plotting our resulting effective
contact angles $\Theta_{\rm max}$ as a function of the substrate
period $a$ for various $\tilde \epsilon$.  The effective angles are
\textit{smaller} than the corresponding Young angles and the
discrepancy grows significantly with increasing hydrophobicity.

We therefore believe that it is reasonable to assume that our
effective contact angles embody the effects of (i) the finite range of
the substrate potential, (ii) finite width of the liquid-vapor
interface, and (iii) line tension.

Interestingly, the details of the geometry seem to play only a minor
role: At fixed hydrophobicity the effective $\Theta_\textrm{max}$ is
practically constant in the case of stripes. Even more importantly,
the pillared geometry yields the {\it same} effective contact angle
for large periods of substrate corrugation ($a \gtrsim 20 \,
\textrm{nm}$), if only the hydrophobicity is the same, validating
assumption (iii). We believe that the deviations for smaller values of
$a$ are due to the effect of the pillar corners, whose relative
importance decreases with increasing corrugation dimensions.

Another remarkable observation is the finding that in the
semi-macroscopic limit ($a \gtrsim 20 \, \textrm{nm}$) the effective
$\Theta_\textrm{max}$ values are fairly similar even for different
hydrophobicities, and deviate much less from each other than the
corresponding Young angles. Even though at these scales the contact
angles are usually considered as macroscopic ones, we emphasize that
the {\it height} of the corrugation, $2 \, \textrm{nm}$, is
nanoscopic, and this may possibly trigger this effect. In our model
hydrophobicity (in other words, surface ``chemistry'') is varied by
modifying $\tilde \epsilon$, i.~e. the energy scale of the attraction,
while $\sigma$, the characteristic length scale affecting the
atomic-scale corrugation, remains unchanged. Apparently, on this level
of description, the phenomenon of super-hydrophobicity is governed
more by geometrical effects and the range of the substrate potential
rather than by the strength of attraction.

\subsection{Considerations on the absence of thermal fluctuations}
\label{ssec:NoThermalNoise}

An important question refers to the extent to which the generality of
our conclusions is affected by the absence of fluctuations in
classical DFT. Above the length scales where microscopic roughness of
the interface sets in, a qualitative estimate regarding the role of
fluctuations can be obtained from capillary-wave
theory~\cite{Smoluchowski, Stillinger} and its application to
fluctuating interfaces~\cite{Werner, Vink, Ismail, Netz}.  According
to this theory, the mean square deviation of the local interface
position from its average location is given by $s^2 =
\frac{k_BT}{2\gamma}\ln\left(\frac{L}{B_{\rm o}}\right)$, where $L$ is
the lateral size of the free interface between corrugations. $B_{\rm
  o}$ stands for a coarse-graining scale~\cite{Werner} below which the
capillary-wave Hamiltonian is not applicable, being of the order of
one nanometer (e.~g. Ref.~\onlinecite{Netz} reported a crossover at
$B_{\rm o} = 0.8 \, \mathrm{nm}$). We thus find that for the
parameters studied in this paper ($L$ of the order of a few tens of
nanometers, and assuming $B_{\rm o} = 1 \, \mathrm{nm}$) the
fluctuation $s$ remains on the sub-nanometer scale. This estimate
suggests that at least long-wave fluctuations should not cause
significant deviations from the observations reported here.

\section{Summary and outlook} 
\label{sec:summ}
%

The applicability of a phenomenological force-balance relation for the
Cassie-Wenzel transition of a water-vapor interface has been studied
by a simple model based upon classical DFT. The method allows to study
arbitrary surface geometries on the nanoscale, where the
phenomenological picture is least obviously valid. In the present
paper we picked corrugations of striped and pillared type.

The macroscopic force-balance relation describes the absolute
stability of the Cassie state in terms of the impalement pressure that
depends on geometrical parameters of the corrugation, the liquid-vapor
interface tension, and the effective contact angle,
$\Theta_\textrm{max}$. An attempt to interpret this angle as the Young
contact angle fails on the nanoscale. Instead, the effective angle
quantitatively satisfying the force-balance is \textit{smaller} than
the corresponding Young's value and takes into account effects of the
finite range of the liquid-solid interactions, line tension and the
diffuse nature of the interface. This suggestion is corroborated by
the fact that in the case of striped geometry the effective contact
angle is independent of the substrate period. Furthermore, for large
periods ($a \gtrsim 20 \, \textrm{nm}$) the effective contact angles
found for striped and pillared geometry are the same at fixed
hydrophobicity, as the influence of the corners of the pillars is not
too large. We do not find any indication that the Cassie-Wenzel
transition can be of {\it second} order at least for some nanoscopic
geometries and hydrophobicity strengths, as was suggested in
Ref.~\onlinecite{NT_MM_slip_2012}. However in that latter study
thermal fluctuations were present and probably played a crucial role
in the transition.

In the present investigation we employed a simple classical DFT model
which captures the long-wavelength properties of a liquid, while
neglecting local-scale packing. The latter can be taken into account
through more elaborate DFT descriptions. We expect that such local
packing will result in a certain shift of the predicted effective
angle in two- and three-dimensional geometries, even if the
macroscopic thermodynamics remains unchanged. However, we also expect
that (at fixed hydrophobicity) these very different geometries will be
still characterized by a unique value of the effective angle, provided
that the corrugation features are not too strongly miniaturized. A
test of this hypothesis is left for future work; within the framework
of the present model this is (at least conceptually) easy to do by
implementing more advanced density functionals. Further possible
extensions include the incorporation of a gas component (air instead
of water vapor) as well as numerical improvements, such as
adaptive-mesh schemes, to reach scales relevant to the range of
optical observations in experiments.

\begin{appendices}

\section{Macroscopic theory of the Cassie-Wenzel transition}
\label{sec:appendix}

\begin{figure}
  \begin{center}
    \includegraphics[width=8.5cm]{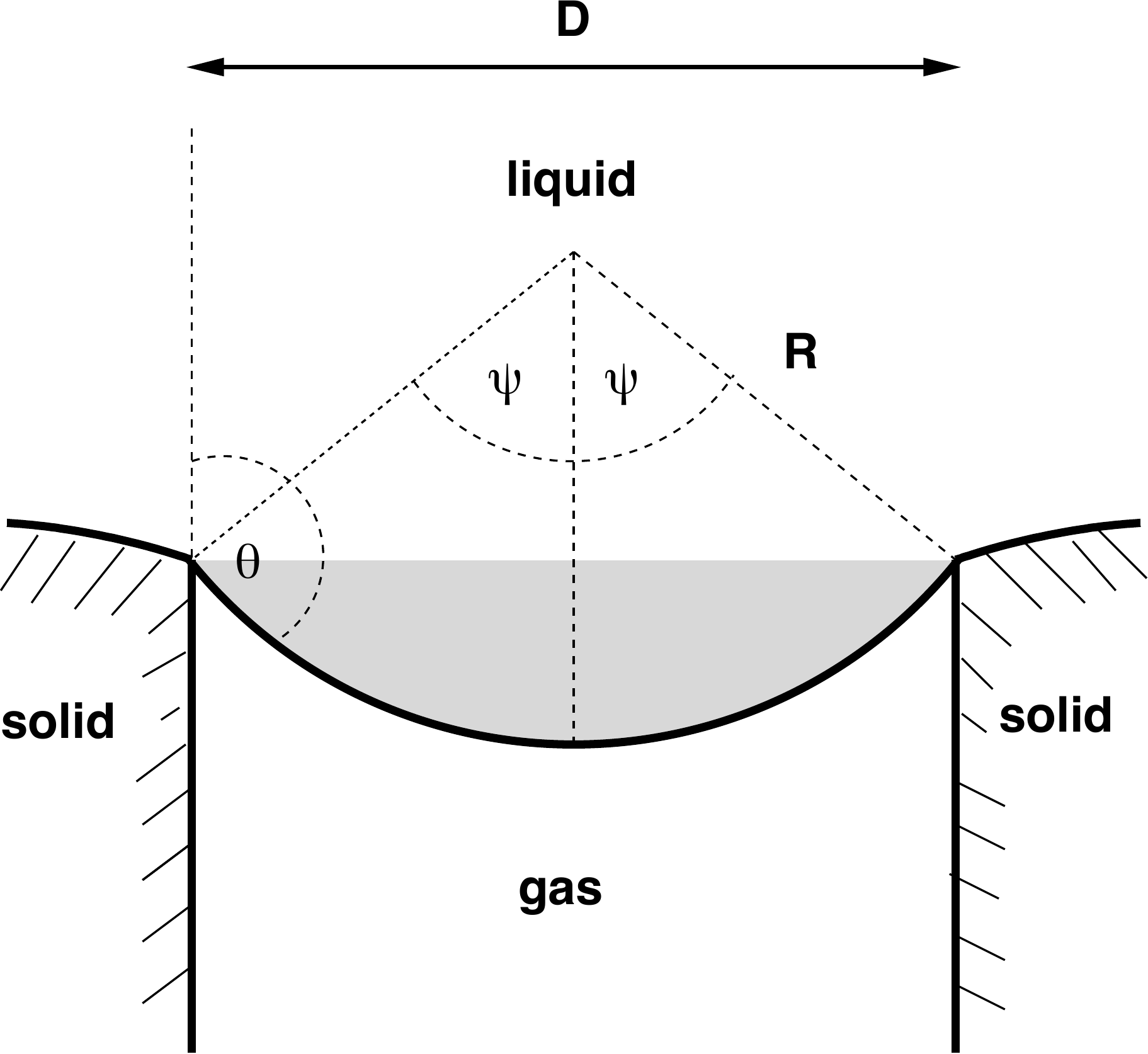}
  \end{center}
  \caption{Simple macroscopic geometry for a liquid on a structured
    surface in the Cassie state.}
  \label{fig:cassiewenzelgeometry1}
\end{figure}

We here derive the critical excess pressure for the Cassie-Wenzel
transition. Figure~\ref{fig:cassiewenzelgeometry1} outlines a simple
geometry that is easy to analyze.

We assume that the situation is translationally invariant in the $y$
direction, i.~e. the direction perpendicular to the drawing plane.  We
assume that the system has an extension $L_y$ in that direction.  We
furthermore assume, for simplicity, that the gas--liquid interface has
the shape of a cylinder surface. Finally, we assume that the
gas--solid interfaces are perpendicular to the (overall) surface, and
that the asperities are so deep that effects of their finite depth
(``sagging'') can be ignored. The figure then shows the liquid
suspended in the Cassie state. We denote the lateral extension of the
cavity with $D$. It is related to the cylinder radius $R$ and the
opening angle $2 \psi$ via
\begin{equation}
  \frac{D}{2} = R \sin \psi ,
\end{equation}
while the effective contact angle $\Theta$ (as defined in the figure)
is given by
\begin{equation}
 \Theta = \psi + \frac{\pi}{2}
\end{equation}
(we use radians as units for angles). The area of the gas--liquid
interface is then given by
\begin{equation}
  A = L_y \, R \, 2 \psi = L_y D \frac{\psi}{\sin \psi} ,
  \label{eq:area}
\end{equation}
while the liquid volume corresponding to the shaded area of the figure
is given by
\begin{eqnarray}
  V_{\rm l} & = &
  L_y \left( R^2 \psi - R \sin \psi \, R \cos \psi \right)
  \nonumber
  \\
  & = &
  \frac{L_y D^2}{4 \sin^2 \psi}
  \left( \psi - \sin \psi \cos \psi \right) .
  \label{eq:volume}
\end{eqnarray}
If we denote the total volume of the cavity with $V_{\rm tot}$, then the
volume occupied by the gas, $V_{\rm g}$, is obviously given by
\begin{equation}
  V_{\rm g} = V_{\rm tot} - V_{\rm l} .
\end{equation}
We now consider the grand potential of the system within the total
volume of the cavity. The bulk contributions are given by
\begin{equation}
  \Omega^{\rm bulk} = - P_{\rm l} V_{\rm l} - P_{\rm g} V_{\rm g} ,
\end{equation}
where $P_{\rm l}$ and $P_{\rm g}$ are the pressures in the liquid and
gas phases, respectively. As a result of the curved interface, these
pressures are not the same (Laplace pressure). We then can define an
excess grand potential (using the gas phase fully occupying the cavity
as a reference state) via
\begin{equation}
  \Omega^{\rm bulk}_{\rm ex} = \Omega^{\rm bulk} + P_{\rm g} V_{\rm tot}
                    = - P_{\rm ex} V_{\rm l} ,
\end{equation}
where the excess pressure is given by
\begin{equation}
  P_{\rm ex} = P_{\rm l} - P_{\rm g} .
\end{equation}
The full excess grand potential has also an interface contribution,
such that
\begin{equation}
  \Omega_{\rm ex} = \gamma_{\rm lg} A - P_{\rm ex} V_{\rm l},
\end{equation}
where $\gamma_{\rm lg}$ is the liquid-gas interface tension.

Normalizing the potential by the factor $D^2 L_y$, we thus obtain
for the normalized potential of the Cassie state
\begin{equation}
  \omega_{\rm C} = \Gamma_{\rm lg} \frac{\psi}{\sin \psi} - P_{\rm ex}
  \frac{1}{4 \sin^2 \psi} \left( \psi - \sin \psi \cos \psi \right) ,
  \label{eq:free_en_C}
\end{equation}
where $\Gamma_{\rm lg} = \gamma_{\rm lg} / D$.

The equilibrium shape of the interface in the Cassie state is then
obtained by minimizing $\omega_{\rm C}$ with respect to $\psi$. After
some straightforward (computer) algebra, this results in
\begin{equation}
  \sin \psi = \frac{P_{\rm ex}}{2 \Gamma_{\rm lg}} ,
\end{equation}
or, returning to the angle $\Theta$,
\begin{equation}
  - \cos \Theta = \frac{P_{\rm ex}}{2 \Gamma_{\rm lg}} . 
  \label{eq:theta}
\end{equation}

\begin{figure}
  \begin{center}
    \includegraphics[width=8.5cm]{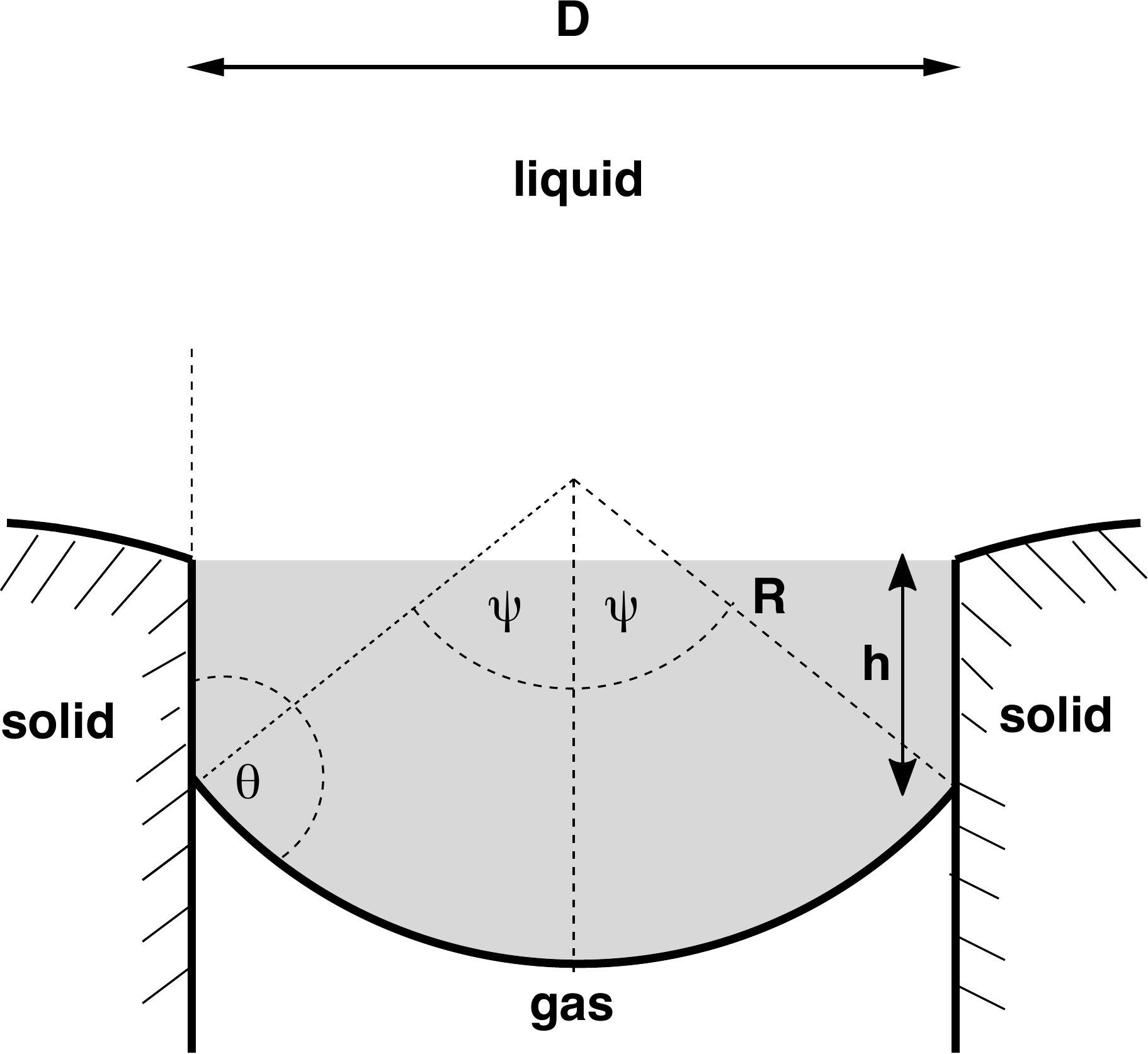}
  \end{center}
  \caption{Macroscopic geometry for a liquid penetrating the cavity.
  The depth of the penetration is $h$.} 
  \label{fig:cassiewenzelgeometry2}
\end{figure}

In order to study the stability of the Cassie state, we compare its
grand potential with the grand potential of a Wenzel-like state, where
the interface is pushed into the cavity by some amount $h$
(Fig.~\ref{fig:cassiewenzelgeometry2}). For this latter potential we
obtain
\begin{eqnarray}
  \Omega_{\rm W} & = & \Omega_{\rm C} - P_{\rm ex} D L_y h
  + 2 \left( \gamma_{\rm sl} - \gamma_{\rm sg} \right) L_y h , \\
  \omega_{\rm W} & = & \omega_{\rm C} -
  \left( P_{\rm ex} - 2 \Delta \Gamma  \right) \frac{h}{D} .
\end{eqnarray}
The second term corresponds to the additional bulk contribution from
the additional liquid that has been pushed into the cavity.
Similarly, the third term is the interface contribution, where
$\gamma_{\rm sl}$ and $\gamma_{\rm sg}$ are the interface tensions of
the solid-liquid and solid-gas interfaces, respectively. The
difference of these values occurs since a solid-gas interface is
replaced with a solid-liquid interface. In the second equation, we
abbreviate $\Delta \Gamma = \left( \gamma_{\rm sl} - \gamma_{\rm sg}
\right) / D$.

The equilibrium configuration is then obtained by optimizing
$\omega_{\rm W}$ with respect to both $\psi$ and $h$. Therefore,
$\psi$ remains unchanged, compared to the Cassie state, while the
solution for $h$ depends on the applied pressure. For $P_{\rm ex} < 2
\Delta \Gamma$ we obtain $h = 0$, i.~e. for small pressures the Cassie
state remains stable. Conversely, for $P_{\rm ex} > 2 \Delta \Gamma$
the solution is $h = \infty$, which corresponds to the Wenzel state.

The critical pressure (or the maximum excess pressure up to
which the Cassie state is stable) is therefore given by
\begin{equation}
  P^{\rm max} = 2 \Delta \Gamma .
\end{equation}
Inserting this into Eq.~(\ref{eq:theta}), we find for the contact angle
at the transition
\begin{equation}
 - \cos \Theta = \frac{\Delta \Gamma}{\Gamma_{\rm lg}}
\end{equation}
or
\begin{equation}
  \gamma_{\rm sg} = \gamma_{\rm sl} + \gamma_{\rm lg} \cos \Theta ,
\end{equation}
which is Young's equation. In other words, the Cassie-Wenzel
transition occurs precisely when the contact angle has reached the
Young value, $\Theta = \tflat$. Therefore, the critical pressure can
also be written as
\begin{equation}
  P^{\rm max} = - 2 \Gamma_{\rm lg} \cos \tflat
  = - \frac{2 \gamma_{\rm lg}}{D} \cos \tflat ,
\end{equation}
which is the formula given in the main text.
  
\end{appendices}

\section*{Acknowledgment} 
\label{sec:ack}
%

We thank Raffaello Potestio for a careful reading of the manuscript,
and his useful comments. This work was partially supported by DFG, SFB
Transregio 146 and ERC for the advanced grant 340391-SUPRO.

\bibliography{microdft_vs_macro}

\end{document}